\documentclass[12pt]{article}
\usepackage{amssymb}
\usepackage{graphicx}
\usepackage{latexsym}
\newcommand{\eq}{\begin{equation}}
\newcommand{\en}{\end{equation}}
\newcommand{\qe}{\end{equation}}
\newcommand{\ear}{\begin{eqnarray}}
\newcommand{\eqa}{\begin{eqnarray}}
\newcommand{\rae}{\end{eqnarray}}
\newcommand{\ena}{\end{eqnarray}}
\newcommand{\beq}{\begin{equation}}
\newcommand{\eeq}{\end{equation}}
\newcommand{\bea}{\begin{eqnarray}}
\newcommand{\eea}{\end{eqnarray}}
\newcommand{\Z}{\mathbb{Z}}
\begin{document}
\begin{titlepage}
\vskip0.5cm
\begin{flushright}
DFTT 08/04\\
SHEP-0405\\
DIAS-STP-04-01 \\
\end{flushright}
\vskip0.5cm
\begin{center}
{\Large\bf  Short distance behaviour of the effective string}
\end{center}
\vskip1.3cm
\centerline{
M. Caselle$^{a}$, M. Hasenbusch$^{b}$
 and M. Panero$^{c}$}
 \vskip1.0cm
 \centerline{\sl  $^a$ Dipartimento di Fisica
 Teorica dell'Universit\`a di Torino and I.N.F.N.,}
 \centerline{\sl via P.Giuria 1, I-10125 Torino, Italy}
 \centerline{\sl
e--mail: \hskip 1cm
 caselle@to.infn.it}
 \vskip0.4 cm
 \centerline{\sl  $^b$  Department of Physics and Astronomy,
                         University of Southampton,}
   \centerline{\sl
                   Southampton, SO17 1BJ,
                        United Kingdom}

 \centerline{\sl
e--mail: \hskip 1cm
 hasenbus@phys.soton.ac.uk}
\vskip0.4 cm
 \centerline{\sl  $^c$ School of Theoretical Physics,
Dublin Institute for Advanced Studies,}
 \centerline{\sl
                  10 Burlington Road, Dublin 4,
                              Ireland}
 \centerline{\sl
e--mail: \hskip 1cm
 panero@stp.dias.ie}
 \vskip1.0cm

\begin{abstract}
We study the Polyakov loop correlator in the (2+1) dimensional
$\Z_2$ gauge model. An 
algorithm that we have presented
recently, allows us to reach high precision results for a large
range of distances and temperatures, giving us the
opportunity  to test predictions of the effective Nambu--Goto
string model. Here we focus on the regime of low temperatures and
small distances. In contrast to the high temperature, large
distance regime, we find that our numerical results are not well
described by the two loop-prediction of the Nambu--Goto model. In
addition we compare our data with those for the SU(2) and SU(3)
gauge models in (2+1) dimensions obtained by other authors. We
generalize the result of L\"uscher and Weisz for a boundary term
in the interquark potential to the finite temperature case.
\end{abstract}
\end{titlepage}

\setcounter{footnote}{0}
\def\thefootnote{\arabic{footnote}}

Recently a renewed interest has been attracted by the effective
string description of the interquark potential in lattice gauge
theories (LGT) \cite{lw01}-\cite{Jahn:2003uz}. On one hand,
in \cite{lw02,chp03} high precision simulations of the interquark 
potential were run,
using algorithms \cite{lw01,chp03,fep2001} that allow 
to measure the Polyakov loop correlation function with highly reduced 
variance.
On the other hand, 
in \cite{kuti03} the lowest excited states in the spectrum of the confining flux tube
in the (2+1) dimensional $\Z_2$ gauge model were studied as well, using a variant of earlier methods. These 
works enabled one to test some longstanding conjectures on the  effective string
description of the interquark potential, and also triggered some
new theoretical effort toward a better understanding of the
effective string model itself. While it seems by now understood
that the first order correction to the interquark potential is
given by the so called L\"uscher term~\cite{lsw}, several other
issues are still open and require further investigation. A
tentative list of these open problems includes:
\begin{itemize}
\item {\bf The determination of higher order terms (and possibly the full
functional form) of the effective string action.}

 The presence of the
 L\"uscher term in the interquark potential only tells us
 that, at  leading order, the effective string action is simply
 a two-dimensional quantum field theory
 of $d-2$ free bosonic fields (one  for each
transverse degree of freedom of the fluctuating string).
 It does not
 help to identify the higher order terms in the action, which should
 describe the string self-interaction. To this end,
 one has to evaluate
 higher order corrections (i.e. higher order powers in $1/R$, $R$ being the
interquark distance) in the interquark potential. We addressed this problem in
 two recent papers \cite{cpp02,chp03},
 testing the simplest possible effective action, i.e. the
 Nambu--Goto string (see below for a detailed discussion) in the large distance,
 finite temperature regime of the 3d Ising gauge model. 
We found an impressive
 agreement between
 Monte Carlo data and the prediction from the
 Nambu--Goto action truncated at second order
\footnote{A similar agreement was
 observed in~\cite{cfghpv} where
  the large distance regime of the interface potential in the 3d Ising model
   was studied},
 but
 apparently no room was left for higher order corrections, which should
 necessarily be present if the Nambu--Goto string is the correct picture.

 \item {\bf The universality issue.}

 As we mentioned above, the first order
 correction to the interquark potential (i.e. the L\"uscher term)
  shows
 an impressive degree of universality and has been detected with the
 precise
  numerical value predicted by the effective string theory
   in all of the models studied up to now (ranging
  from the 3d gauge Ising model to the SU(2) and SU(3) models,
  both in d=3 and in
  d=4). Which is the situation for higher order corrections? Do they also show
  the same universal behaviour? Preliminary results
  suggest that this is not the case \cite{lw02}, however these evidences were
  obtained in the short distance regime only (at present only in the case of the
  3d Ising model one can reach the large distance regime),
  where  non-universal boundary terms are present (see last item below)
  and make the
  analysis much more difficult.

 \item{\bf The boundary corrections.}

 As it was recently pointed out by L\"uscher and Weisz, the
 presence of two boundaries (the two Polyakov loops) in the
 finite temperature geometry
 makes it necessary to
 include in the possible higher order terms of the string action also
 contributions localized at these boundary
 (which we shall call in the following  ``boundary terms'')
 which are proportional to a
 non-universal parameter, which we shall denote as $b$ in the following.
 These terms induce
 corrections to the expected form of the interquark potential
 (which we shall
 call ``boundary corrections'' in the following ):
 these corrections are proportional to $b$. The open
 problem in this case is to evaluate $b$ in the various LGT's, to
 check if it can
 be compatible with zero and more generally if it depends on some relevant
 feature of the underlying LGT. To this end, it would be very
 useful to derive
 the whole functional form (i.e. the explicit dependence from $R$ and $L$)
 of the boundary correction.

 \item{\bf High precision test of the L\"uscher correction.}

 While, as we mentioned above,
 it is by now accepted  that the first order
correction to the interquark potential is given by the L\"uscher
term, a high precision test of this result in the finite
temperature regime is still missing. It would be interesting to
have a quantitative estimate both of the range of values of the
interquark distance in which the L\"uscher term well describes the
interquark potential and of the numerical uncertainty in the
determination of its value. This is particularly important in view
of the recent results~\cite{lw02,ns01} which show that the bosonic
string seems to work at surprisingly short distances.

 \item{\bf A ``Casimir energy paradox''?}
 
Finally, we mention the fact that in \cite{kuti03} the authors 
presented numerical results showing that in a distance range (below 1 fm)
where the L\"uscher term
in the interquark ground state potential is already observed, 
they only found a few stable excitations of the confining flux, 
whose spectrum turned out to be  
grossly distorted compared with the string prediction. 

\end{itemize}

The aim of this paper is to address 
some of
these issues by looking once again
 at the 3d gauge Ising model, which allows to perform
 very  precise simulations while keeping all of the possible 
sources of systematic
 error under control. To this end, we performed a set of new
 simulations in the  {\bf short distance, low temperature}  
regime of the interquark potential, which we
 used first to make a high precision test of the L\"uscher term (as discussed
 above)  and then to study the higher order terms of the effective string
 action. In this respect the present paper can be considered as the completion of our
 previous paper~\cite{chp03} in which we performed
a similar study in the 
{\bf large
 distance, high temperature} regime of the interquark potential. 
Our results can also be compared with 
the findings of an analogous study, recently published 
in \cite{kuti03}. A preliminary account of our results can be found
in~\cite{chp_proc}. A non-trivial problem
 one has to face when looking at the higher order
 effective string terms is the presence of the
 non-universal boundary
 correction discussed above. In order to deal with this problem
 we evaluated
 the functional form of the boundary correction (a result which is rather
 interesting in
 itself as we shall see below),
 in the framework of the zeta-function regularization. In this
 way, we
may predict the large distance behaviour of the boundary
correction, and use the high precision results of~\cite{chp03} to
fix the non-universal constant in front of it. As we shall see,
our data strongly suggest that this value is compatible with zero
for the 3d gauge Ising model.

 A side consequence of this analysis
is that we shall be able to show in a rather unambiguous way that,
at least in the short distance regime the interquark potential, in
the 3d Ising model cannot be described by a simple Nambu--Goto
string. We shall then conclude the paper with a tentative
comparison of our results with the short distance behaviour of
other LGT's trying to shed some light on the universality issue
discussed above.

This paper is organized as follows. We start in sect.~\ref{sect1}
with a short summary of the effective string description of the
interquark potential.
  Then in the following two sections we compare our new
  Monte Carlo results with the
  effective string predictions in the short distance regime.
  In sect.~\ref{sectnew}, we discuss the first order correction
(the ``L\"uscher'' term), while in sect.~\ref{sect3} we address
the issue of the higher order corrections. Sect.~\ref{boundary} is
then devoted to the study of
    the functional form of the boundary correction.
These results are used in sect.~\ref{sect2} to study the boundary
correction in the large distance regime. Sect.~\ref{sect4} is
devoted to some concluding remark and to the comparison of our
Ising results with those obtained in the SU(2) and SU(3) gauge
theories in $d=3$.
 Finally, the detailed derivation of the functional
  form used in sect.~\ref{boundary} is reported in the Appendix.

\section{Summary of known theoretical results on the effective string description of
the interquark potential}
\label{sect1}

We refer to our previous papers~\cite{cpp02,chp03} for a
detailed discussion about the effective
string picture and its peculiar realization in the finite temperature
geometry (i.e. in the case of Polyakov loop correlators). Here,
we only recall some basic facts (more details can be found in the appendix).
\begin{itemize}

\item
In  finite temperature LGT's the
interquark potential can be extracted by looking at  correlators
of Polyakov loops in the confined phase.
The correlator of two loops $P(x)$  at
a distance $R$ and at a temperature $T=1/L$ ($L$ being
the size of the lattice
in the compactified ``time'' direction) is given by

\eq
G(R)\equiv
\langle P(x)P^\dagger(x+R) \rangle \equiv {\rm e}^{-F(R,L)}~~~.
\label{polya}
\en
\noindent
The free energy $F(R,L)$ is expected to be described, as a
first approximation,  by the so called ``area law''
\eq
F(R,L)\sim \sigma L R + k(L)
~~~,\label{area}
\en
where $\sigma$ denotes the string tension
and $k(L)$ is a non-universal
 constant depending only on $L$.

\item According to the effective string picture, in the rough
phase of the theory one must add to the area law of
eq.~(\ref{area}) a correction due to quantum fluctuations of the
flux tube (``effective string corrections''). Such a correction is
expected to be a complicated function of $R$ and $L$, but if one
neglects the contributions due to the string self-interaction
terms and from the boundary terms which we shall discuss below,
then one finds \eq F(R,L)\sim  F_q(R,L)=\sigma L R + k(L)+
F^1_q(R,L) ~~~\label{a+q} \en with \eq
F_q^1(R,L)=(d-2)\log\left({\eta(\tau)}\right) \hskip0.5cm
;\hskip0.5cm {-i}\tau={L\over 2R}~~~, \label{bos} \en \noindent
where $(d-2)$ is the number of transverse dimensions, $\eta$
denotes the Dedekind eta function \eq
\eta(\tau)=q^{1\over24}\prod_{n=1}^\infty(1-q^n)\hskip0.5cm
;\hskip0.5cmq=e^{2\pi i\tau}~~~,\label{etafun} \en and the labels
$q$ and $1$ in $F^1_q$ recall that this is the {\bf first order}
 term in the expansion of the {\bf quantum} fluctuations of the
flux tube. Eq.~(\ref{a+q}) is referred to as the ``free string
 approximation''.
\vskip0.3cm

\item The Dedekind function has two different expansions
--- both of which turn out to be convergent for any finite value of $L$ and $R$ ---
 which are most suitable for the two regimes $2R<L$ and
$2R>L$, respectively. These expansions
 are related to each other by the so called ``modular
transformation".

\begin{description}
\item{For $2R<L$}
\eq
F_q^1(R,L)=\left[-\frac{\pi L}{24 R}
+\sum_{n=1}^\infty \log (1-e^{-\pi nL/R})\right](d-2)~~~.
\label{zsmalltot}
\en

The first term of this expansion is the well known ``L\"uscher
term'' ~\cite{lsw}.

\item{For $2R>L$}
\eq
F_q^1(R,L)=\left[-\frac{\pi R}{6 L}+\frac{1}{2} \log\frac{2R}{L}
+\sum_{n=1}^\infty \log (1-e^{-4\pi nR/L})\right](d-2)~~~.
\label{zbigtot}
\en

In this case, the first term of the expansion is
proportional to $R$ and acts to lower the
string tension.

\end{description}

Unless we are in the intermediate region $R\sim L/2$,
 the exponentially decreasing
terms which appear in eq.s~(\ref{zsmalltot})
and (\ref{zbigtot}) can be neglected. We shall largely use this approximation in
the following.

\item At large enough temperatures, i.e. for small values of $L$,
or small enough values of $R$, higher order terms in the expansion
of the flux tube quantum fluctuations become important and cannot
be neglected. These terms encode the string self-interaction and
 depend on the particular choice of the effective string action. The simplest
 proposal (discussed in~\cite{cpp02,chp03}) is the Nambu--Goto
 action, in which the string configurations are simply weighted
 by the area of the world-sheet. Its contribution to the free energy turns out
 to be (fixing for simplicity $d=3$) (see \cite{df83} and \cite{chp03})
\eq F_q^{2}(R,L)=-\frac{\pi^2 L}{1152\ \sigma R^3}\left[2
E_4(\tau)-E_2^2(\tau)\right] \;\;, \label{nlo} \en

where $E_2$ and $E_4$ are the Eisenstein functions. The latter can be
expressed in power series
\eqa
E_2(\tau)&=&1-24\sum_{n=1}^\infty \sigma(n) q^n\\
E_4(\tau)&=&1+240\sum_{n=1}^\infty \sigma_3(n) q^n\\
q&\equiv& e^{2\pi i\tau} \;\;,
\ena
where $\sigma(n)$ and $\sigma_3(n)$ are, respectively, the sum of all
divisors of $n$ (including 1 and $n$), and the sum of their cubes.

\end{itemize}

\section{High precision test of the L\"uscher term}
\label{sectnew}
\subsection{The simulations}
In order to perform a high precision test of the L\"uscher term we
run a set of Monte Carlo simulations in the short distance regime.
In tab.~\ref{tabnew1} we report some information on our
simulations. The choice of very large lattice sizes in the
``temperature'' direction (i.e. the direction of the Polyakov
loops)
 ensures us that we are in the very low temperature domain. The temperature of our
 present simulations
ranges from $T/T_c=1/20$ for $L=80$ at $\beta=0.73107$ up to
$T/T_c=1/5$ for $L=20$ at $\beta=0.73107$. The values for the
$1/L$ corrections estimated in~\cite{chp03} indicate that for
$T\leq\frac{T_c}{10}$ the contribution of possible higher order
effective string terms due to the finiteness of the lattice size
in the temperature direction is completely negligible within the
precision of our data while the data at $T=T_c/5$ are at the
border of our precision. Thus, at least for $T\leq\frac{T_c}{10}$
 we may neglect
possible $1/L$ corrections and we can concentrate on the $1/R$
corrections only. Notice that this is exactly the opposite
situation with respect to our previous paper~\cite{chp03} in which
the temperature was much higher and the $1/L$ corrections were
dominating. The sample at the smallest value of $\beta$:
$\beta=0.65608$ is characterized by a rather small correlation
length and we should therefore expect large scaling violations.

For a detailed discussion of the algorithm that we used, of the
simulation setting and more generally on the 3d gauge Ising model,
we refer the reader to our previous paper~\cite{chp03}. The
results of our simulations are reported in tab.~\ref{tabnew2} and
in tab.~\ref{tab1} \footnote{Notice  that also the result at
$R=80$ listed in tab.~\ref{tab1} and those at $R=40$ listed in
tab.~\ref{tabnew2} are meaningful (even if $80 > 128/2$ and $40 >
64/2$) due to the non-trivial mapping of the boundary conditions
from the Ising  gauge to the Ising spin model under duality
transformation (see~\cite{chpz02}, Sect. 4.3 for a detailed
discussion of this point). In particular, performing our
simulations in the 3d Ising spin model (as we did), we neglected
the anti-periodic boundary conditions, which would produce the
``echo'' contribution due to the distances (128-80) and (64-40)
respectively.}.

\begin{table}[h]
\begin{center}
\begin{tabular}{|c|c|c|r|l|l|r|}
\hline
\multicolumn{1}{|c}{$\beta$}
&\multicolumn{1}{|c}{$L_{c}\equiv 1/T_c$}
&\multicolumn{1}{|c}{$L$}
&\multicolumn{1}{|c}{$N_s$}
&\multicolumn{1}{|c}{$\xi$}
&\multicolumn{1}{|c}{$\sigma$}
&\multicolumn{1}{|c|}{$R_c$} \\
\hline
0.65608 & 2  &20& 32 & 0.644(1) &0.20487(1) & 2.71\\
\hline
0.73107 & 4  &20,40,80& 64 & 1.456(3)&0.0440(3) & 5.84\\
\hline
0.75180 &  8  & 80 & 128 &3.09(1)&0.010560(18) & 11.92\\
\hline
\end{tabular}
\end{center}
\caption {\sl A few information on our simulations. In the first
column the value of $\beta$ is given and in the second the inverse
of the critical temperature. The third and fourth columns contain
the values of lattice size in the ``temperature'' direction $L$
and in the two spacelike directions $N_s$. In the last three
columns the values of the exponential correlation length (obtained
by interpolating the values reported in~\cite{Acch97}
and~\cite{Ch97} and by the analysis of the low temperature
series~\cite{arisue}), the value of the string tension in
dimensionless units, and the value of $R_c$ defined as $R_c\equiv
\sqrt{1.5/\sigma}$ (see sect.~2.7 of ref.~\cite{chp03}) are
provided.} \label{tabnew1}
\end{table}

\begin{table}[h]
\begin{center}
\begin{tabular}{|r|l|l|l|}
\hline
\multicolumn{1}{|c}{$R$}
&\multicolumn{1}{|c}{$L=20$}
&\multicolumn{1}{|c}{$L=40$}
&\multicolumn{1}{|c|}{$L=80$} \\
\hline
  4&  0.370163(58)&  0.136976(25)&  0.0187711(45)\\
  5&  0.382971(60)&  0.146536(27)&  0.0214749(53)\\
  6&  0.391093(62)&  0.152723(31)&  0.0233307(56)\\
  7&  0.396557(66)&  0.157044(31)&  0.0246517(61)\\
  8&  0.400670(68)&  0.160041(33)&  0.0256028(68)\\
  9&  0.403491(72)&  0.162210(33)&  0.0263272(71)\\
 10&  0.405866(72)&  0.163930(36)&  0.0268669(76)\\
 11&  0.407682(72)&  0.165229(36)&  0.0272785(81)\\
 12&  0.409213(77)&  0.166123(36)&  0.0276178(83)\\
 13&  0.410384(82)&  0.166999(38)&  0.0278800(84)\\
 14&  0.411507(81)&  0.167647(38)&  0.0280970(86)\\
 15&  0.412345(87)&  0.168165(41)&  0.0282665(88)\\
 16&  0.413234(86)&  0.168697(41)&  0.0284264(93)\\
 18&  0.414431(90)&  0.169435(43)&  0.0286463(95)\\
 20&  0.415658(95)&  0.169883(43)&  0.0288094(97)\\
 30&  0.418863(110)& 0.171406(48)&  0.0292359(109)\\
 40&  0.420810(124)& 0.171981(50)&  0.0293721(109)\\
\hline
\end{tabular}
\end{center}
\caption{\sl Values of the ratio between two successive
Polyakov loop correlators
$G(R+1)/G(R)$ for
various values of $R$ and $L$, at $\beta=0.73107$~.}\label{tabnew2}
\end{table}

\begin{table}[h]
\begin{center}
\begin{tabular}{|r|l|}
\hline
\multicolumn{1}{|c}{$R$}
&\multicolumn{1}{|c|}{$L=80$}
 \\
\hline
 8 &  0.382613(58)  \\
10  & 0.396313(61)  \\
12 &  0.405088(65)  \\
14  & 0.411017(69)  \\
16  & 0.415268(71)  \\
18  & 0.418048(74)  \\
20  & 0.420344(74)  \\
22  & 0.422062(78)  \\
24  & 0.423323(78)  \\
26  & 0.424486(78)  \\
28  & 0.425496(81)  \\
30  & 0.425958(81)  \\
32  & 0.426747(82)  \\
36  & 0.427513(83)  \\
40  & 0.428255(84)  \\
60  & 0.430097(87)  \\
80  & 0.431068(88)  \\
\hline
\end{tabular}
\end{center}
\caption
{\sl Same as tab.\ref{tabnew2} but for the data at $\beta=0.75180$~.}
\label{tab1}
\end{table}

\begin{table}[h]
\begin{center}
\begin{tabular}{|r|l|}
\hline
\multicolumn{1}{|c}{$R$}
&\multicolumn{1}{|c|}{$L=20$}
 \\
\hline
 2 &  0.010583(3)  \\
 3 &  0.013032(4)  \\
 4 &  0.014370(4)  \\
 5 &  0.015101(4)  \\
 6 &  0.015540(4)  \\
 7 &  0.015817(4)  \\
 8 &  0.016002(4)  \\
 9 &  0.016151(5)  \\
10 &  0.016240(5)  \\
12 &  0.016383(5)  \\
14 &  0.016476(5)  \\
16 &  0.016549(5)  \\
18 &  0.016608(5)  \\
20 &  0.016643(6)  \\
\hline
\end{tabular}
\end{center}
\caption
{\sl Same as tab.\ref{tabnew2} but for the data at $\beta=0.65608$~.}
\label{tabl2}
\end{table}

\subsection{Test of the L\"uscher term}
We performed our analysis in three steps. First we made, following
what is usually done in the literature, a rather naive test by
directly fitting the data with a $1/R$ correction. As we shall see
below, this choice is correct only in a very narrow range of
interquark distances, since for large enough values of $R$ the
exponentially decreasing terms $\exp(-\pi n L/R)$ contained in
eq.~(\ref{zsmalltot}) start to matter and cannot be neglected. In
the second level of analysis we fit the data with the whole
functional form of the free bosonic correction,
eq.~(\ref{zsmalltot}), finding a remarkable agreement between the
data and our theoretical expectations. Finally in the third level
of analysis we combine the data in such a way that the string
tension is eliminated as parameter. Thus, we can make an absolute
comparison, with no free parameter to fit, between our data and
the effective string prediction.
 Before going into the details of the
analysis  let us stress a point which will be important in the following.
Due to the algorithm we used, each number in
 tab.s~\ref{tabnew2} and  \ref{tab1} corresponds to a
 different simulation. Hence the sets of data
 we fit are completely uncorrelated and we can safely
 trust both on the $\chi^2$
 obtained from the fit and on the best fit results for the parameters.

\subsubsection{Fitting with a $1/R$ correction}
We fitted the data reported in tab.~\ref{tabnew2} and
tab.~\ref{tab1} according to the law \eq
-\frac1L\log\left(\frac{G(R+1)}{G(R)}\right)=a-b\left(\frac{1}{R+1}-\frac1R\right)~~~.
\label{eqfitnew} \en With this normalization the bosonic string
model predicts $a=\sigma$ and $b=\pi/24=0.13089...$ .

If we fit, for a given choice of $\beta$ and $L$, all the values
of $R$ listed in the tables,
 the reduced $\chi^2$ turns out to be very high and the best fit values of
$a$ and $b$ are very far from the expected values. This indicates
 that at short $R$
we have higher order (string) corrections which are proportional
to higher powers of
 $1/R$ and compromise
 the fit.
The standard way to deal with this type of
  behaviour is to repeat the fit eliminating the data one after the other
   starting from those with the lowest values of
  $R$, until
for some value of $R_{min}$ an acceptable (i.e. order unity) reduced $\chi^2$
is finally reached.
  However it turns out that for all our samples, except the one at $T=T_c/20$ this
  scenario never occurs. Instead we see that after a rapid decrease, the $\chi^2_r$
  starts again to increase as $R_{min}$ is increased and the minimum of this shape
  never goes below 1. As an example, we report in tab.~\ref{tabfit1} the results of the
  fit for the data at $\beta=0.73107$ and $L=40$.
The reason for this behaviour is that
  corrections for {\bf large} values of $R$ are present in the data. These corrections
  are nothing but
  the exponentially decreasing terms of eq.~(\ref{zsmalltot}).
  If we neglect them in the fit, we shall never be able to extract the correct value of
 the L\"uscher term, as the best fit values reported in tab.~\ref{tabfit1} clearly
  show. If the temperature is low enough, there is a range of values of $R$ in which
  both the small $R$ and the large $R$ corrections become negligible and the expected
  value of the L\"uscher term can be recovered. This range can be found by iteratively
  eliminating the smallest and the largest values of $R$ in the fit and
{\bf
 looking for a
 stable and acceptable  value of $\chi^2_r$}.

However, if $T$ is not small enough, this procedure does not
converge. This is the case for instance of our data at
$\beta=0.73107$ and $L=20$. We report in tab.~\ref{tabfit2} the
best fit results
  obtained in this way for the other three samples.

\begin{table}[h]
\begin{center}
\begin{tabular}{|r|r|l|l|}
\hline
\multicolumn{1}{|c}{$R_{min}$}
&\multicolumn{1}{|c}{$\chi^2_r$}
&\multicolumn{1}{|c}{$\sigma$}
&\multicolumn{1}{|c|}{$b$} \\
\hline
 4&  300\phantom{.0}&   0.044123(2)&  0.1147(1)\\
 5&  108\phantom{.0}&   0.044067(2)&  0.1211(2)\\
 6&  48\phantom{.0}&    0.044033(2)&  0.1260(2)\\
 7&  31\phantom{.0}&    0.044012(3)&  0.1295(3)\\
 8&  21\phantom{.0}&    0.043995(3)&  0.1329(4)\\
 9&  15\phantom{.0}&    0.043981(4)&  0.1361(6)\\
 10&  13.2            & 0.043970(4)&  0.1390(7)\\
 11&  9.9            &  0.043958(4)&  0.1426(9)\\
 12&  5.4            &  0.043945(5)&  0.1470(11)\\
 13&  5.1            &  0.043939(5)&  0.1492(14)\\
 14&  3.9            &  0.043931(6)&  0.1525(17)\\
 15&  3.4            &  0.043925(6)&  0.1555(21)\\
 16&  4.2            &  0.043924(7)&  0.1570(26)\\
\hline
\end{tabular}
\end{center}
\caption{\sl Results of the fit according to
 eq.~(\ref{eqfitnew}) for the data at
$\beta=0.73107$ and $L=40$.
In the first column the minimum value of $R$ included in
the fit is given.
The second column contains the reduced $\chi^2$.
In the last two columns the best
fit results for $\sigma$ and for the L\"uscher term are given, respectively.}
\label{tabfit1}
\end{table}

\begin{table}[h]
\begin{center}
\begin{tabular}{|c|c|c|c|c|c|c|}
\hline
$\beta$& $L$& $T/T_c$ &$R$ & $\chi^2_r$ & $\sigma$ & $b$   \\
\hline
0.73107&80&1/20 & 10-20&  0.6&   0.044030(4)\phantom{0}&0.1304(6)\phantom{0}\\
0.73107&40&1/10 & \phantom{0}9-13&  1.5&   0.044042(10)&  0.1285(12)\\
0.75180&80&1/10 & 18-26&  0.4&   0.010532(4)\phantom{0}&  0.1264(19)\\
\hline
\end{tabular}
\end{center}
\caption{\sl Results of the fit according to eq.~(\ref{eqfitnew}).
In the first three columns the values of $\beta$ and $L$ of the data used
in the fit and the corresponding value of $T/T_c$ are provided.
In the fourth column, we  give the range of values of $R$
included in the fit. In the fifth column the reduced $\chi^2$ is given.
The last two columns contain the best
fit results for $\sigma$ and for the L\"uscher term.}
\label{tabfit2}
\end{table}

We insisted in this type of fit because it is widely used in the literature
and, as we have seen, it can easily lead to misleading conclusions.
In particular one should be aware that:
\begin{itemize}
\item if $L$ is not large enough,
i.e. if the temperature is not small enough, the
range of interquark distances within which  this type of fit can be performed
vanishes. The precise threshold in temperature depends on the precision of the data.
In our case the threshold was around $T\sim T_c/10$.
\item
 If one nevertheless performs the fit outside the allowed region  the reduced $\chi^2$
 becomes larger than 1 and the values of the L\"uscher term which one extracts from
 the fit turns out to be (erroneously) different from the expected
 $\pi/24$ factor.
\item If one enforces the expected $\pi/24$ coefficient and performs the fit outside
the allowed range then the values of the string tension extracted from the fit may
show relevant systematic deviations from the correct value.
\end{itemize}
We think that these observations should be carefully
taken into account when dealing with the
interquark potential extracted from Polyakov loops in any LGT, i.e. also in the
physically more interesting SU(2) and SU(3) cases, if the precision reached
by the simulations is high enough.

\subsubsection{Fitting with the whole functional form of the bosonic string
correction}
In order to keep into account the next to leading terms in eq.~(\ref{zsmalltot}) we
 fitted the data reported in tab.~\ref{tabnew2} and tab.~\ref{tab1} according to the
law \eq
-\frac1L\log\left(\frac{G(R+1)}{G(R)}\right)=a+\frac{c}{L}\left(F_q^1(R+1,L)-
F_q^1(R,L)\right)~~~. \label{eqfitnew2} \en With this
normalization $a$ is again the string tension while $c=1$ means
that we have a perfect agreement between the whole bosonic string
prediction (L\"uscher term plus exponentially decreasing
corrections) and the data. In tab.~\ref{tabfit3} we report as an
example the result of the fits in the particular case
$\beta=0.75180$, $L=80$. Similar results are obtained with the
other values of $\beta$ and $L$.

\begin{table}[h]
\begin{center}
\begin{tabular}{|r|r|l|l|}
\hline
\multicolumn{1}{|c}{$R_{min}$}
&\multicolumn{1}{|c}{$\chi^2_r$}
&\multicolumn{1}{|c}{$\sigma$}
&\multicolumn{1}{|c|}{$c$} \\
\hline
 8&  181\phantom{.0}&   0.0105622(7)&  0.824(1)\\
10&   47\phantom{.0}&   0.0105454(8)&  0.883(2)\\
12&   15\phantom{.0}&   0.0105364(9)&  0.922(2)\\
14&   6.6&   0.0105315(10)&  0.949(3)\\
16&   3.9&   0.0105285(11)&  0.967(5)\\
18&   1.2&   0.0105255(12)&  0.990(6)\\
20&   1.1&   0.0105247(14)&  0.996(8)\\
22&   1.0&   0.0105241(15)&  1.003(10)\\
\hline
\end{tabular}
\end{center}
\caption{\sl Results of the fit according to eq.~(\ref{eqfitnew2})
for the data at $\beta=0.75180$, $L=80$. The first column contains
the minimum value of $R$ included in the fit (recall that for this
value of $\beta$ the correlation length is twice the one of
$\beta=0.73107$ and a similar doubling of all the length scales
must be kept into account when comparing the data at the two
values of $\beta$). In the second column the reduced $\chi^2$ is
given. In the last two columns we provide the best fit results for
$\sigma$ and for the coefficient in front of the bosonic string
correction.} \label{tabfit3}
\end{table}

We find
a very good agreement between our data and the bosonic effective
string prediction,
{\sl keeping in the fit all the interquark distances up to $R=80$}.
 The uncertainty in
our determination of $c$ is less than $1\%$. The decrease in the
reduced $\chi^2$ indicates that the higher order string
contributions can be neglected for $R\geq 20$. Looking at the
string tension results, we see that the best fit value changes by
a rather large amount (with respect to the quoted errors) as
$R_{min}$ increases. This observation has two relevant
implications:
\begin{itemize}
\item As we mentioned above, a large systematic deviation in the
best fit estimates of the string tension has to be expected,  if
the effective string corrections are not properly taken into
account. \item Even if the first order effective string correction
is properly taken into account, systematic deviations in the
string tension may still appear if too small values of $R$ are
included in the fits.
\end{itemize}
Our results for the other values of $\beta$
 are summarized in tab.~\ref{tabsigma}. Looking at this table one can see that
the data do not show scaling
violations and that the expected value $c=1$ is found even 
for $\beta=0.65608$ (for which instead
any bulk observable
shows very large scaling violations)~!
This represents a further nontrivial test of the whole picture, since the effective string picture
is indeed expected to hold in the whole range
$\beta_r <\beta< \beta_c$ ($\beta_r$ and $\beta_c$ being respectively the roughening  and the
deconfinement critical points of the gauge model) and should abruptly break
down for $\beta<\beta_r$. It is important to recall that in the above fits we neglected
the short distance data. As we shall see below the short distance deviations from the free string
behaviour are instead affected by scaling corrections.

In the following section we shall make a further step in our analysis. We shall perform
 some kind of ``absolute
test'' of the effective string, combining the data so as to
eliminate the  string tension from the game as well, thus only
leaving the effective string corrections in the data.

\subsubsection{An
``absolute" test of the effective string picture} 
\begin{figure}
\centering
\includegraphics[height=12cm]{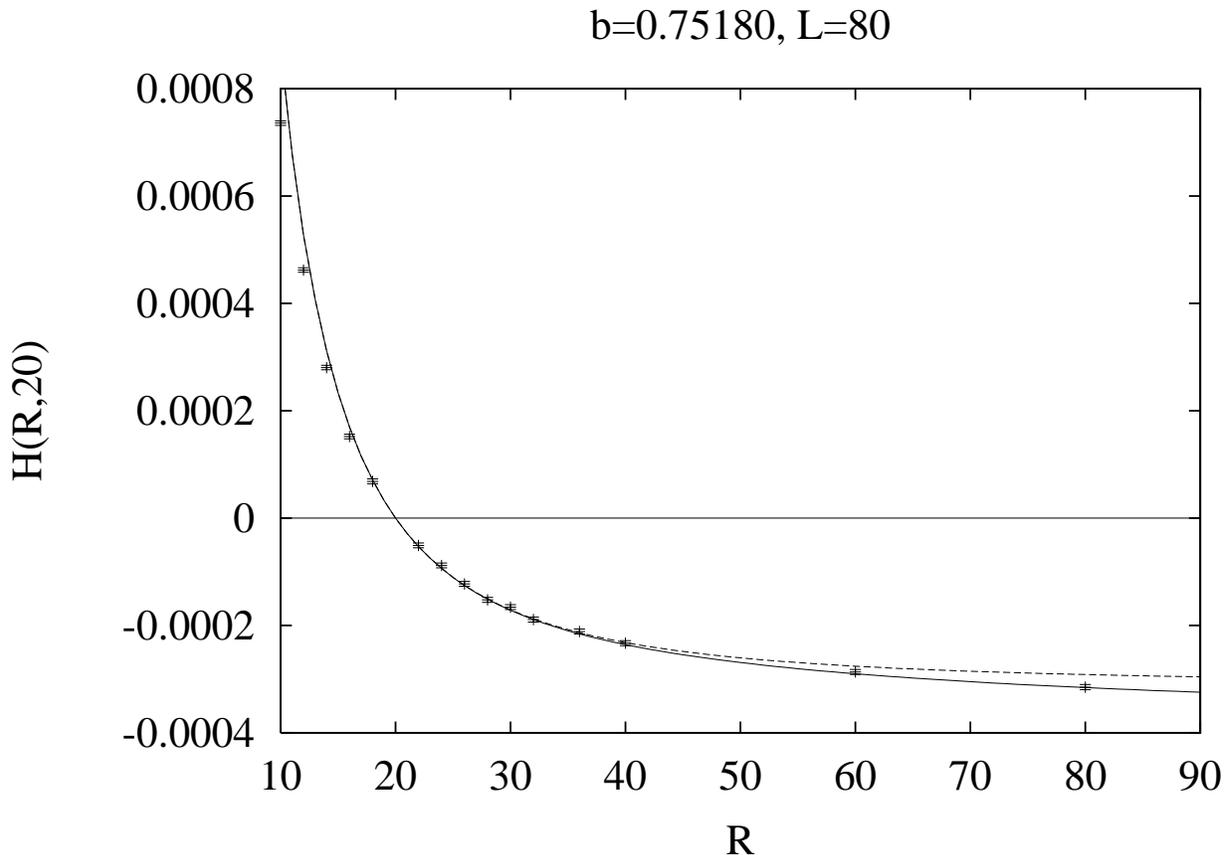}
\caption{Comparison between our values of $H(R,20)$, the bosonic
string prediction, eq.~(\ref{zsmalltot}) (continuous line) and the
$1/R$ approximation to eq.~(\ref{zsmalltot}) (dashed line) for the
data at $\beta=0.75180$, $L=80$.} \label{bnew1}
\end{figure}

\begin{figure}
\centering
\includegraphics[height=12cm]{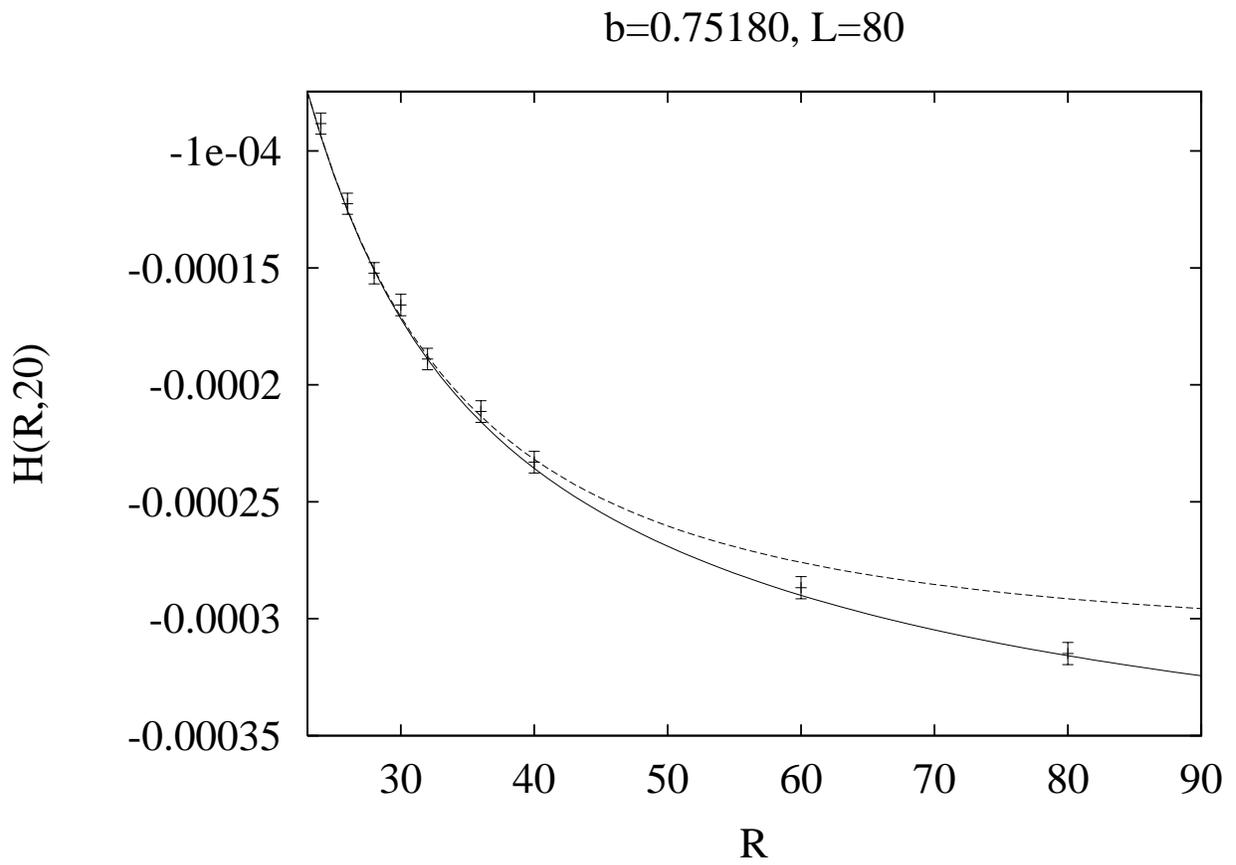}
\caption{Same as fig.~\ref{bnew1}, but with a larger resolution.}
\label{bnew2}
\end{figure}

This can be
easily done by building the combinations \eq H(R_1,R_2)\equiv
-\frac1L\log\left(\frac{G(R_1+1)G(R_2)}{G(R_1)G(R_2+1)}\right) \en
in which the string tension disappears. We plot in
fig.~\ref{bnew1} and \ref{bnew2} a sample of the results we
obtain. In particular we concentrate on the sample at the largest
value of $\beta$ and fix one of the two entries to the first
acceptable
 value of
$R_{min}$ as extracted from the previous analysis, i.e. $R_2=20$.
In fig.s~\ref{bnew1} and \ref{bnew2} we compare our results with
the bosonic string
 prediction for $H(R,20)$ obtained using eq.~(\ref{zsmalltot}) and also with what one
 would obtain (dashed lines in fig.s~\ref{bnew1} and \ref{bnew2}) keeping only the
 $1/R$ correction in eq.~(\ref{zsmalltot}). We find for $R_1>20$ an impressive
 agreement between our data and the effective string predictions. All the data agree
 within the errors. Instead, for $R_1<20$ the large deviations from the free effective
 string appear. We shall discuss these deviations in the forthcoming sections.
 It is very interesting to observe that,
 in agreement with the results
 of the first level analysis discussed above,  for large enough values of $R$
 the contribution due to the
 next to leading, exponentially decreasing, terms of eq.~(\ref{zsmalltot}) is clearly
 visible in the data. It is clear that these terms cannot be neglected in the fits.

\begin{figure}
\centering
\includegraphics[height=12cm]{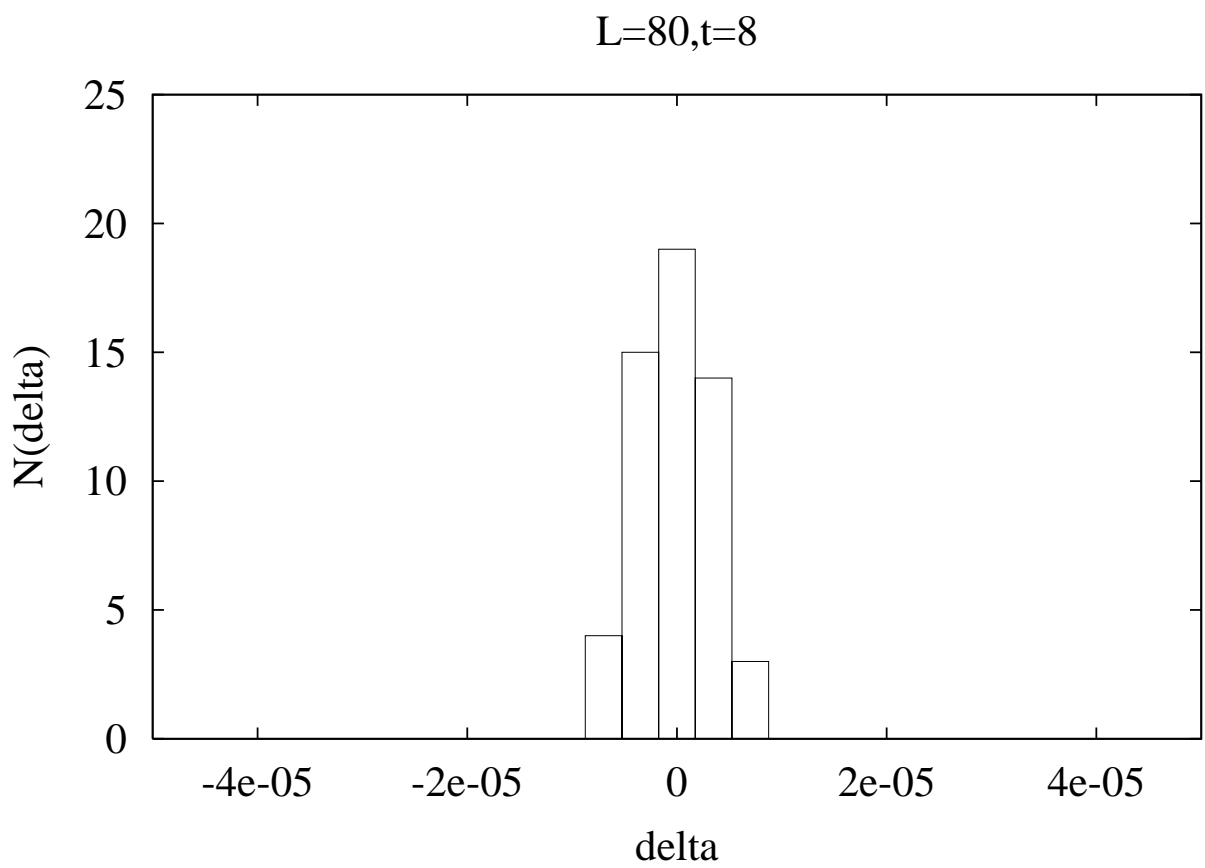}
\caption{Histogram of $\Delta(R_1,R_2)$ for all the values of $R_1,R_2\geq 20$.
Notice the very small scale of the $\Delta$ axis.}
\label{isto80t8}
\end{figure}

Finally in order to have a global view of all our data we built
the differences
 \eq
\Delta(R_1,R_2)\equiv
H(R_1,R_2)-\frac1L\left(F_q^1(R_1+1,L)+F_q^1(R_2,L)
-F_q^1(R_2+1,L)-F_q^1(R_1,L)\right)~~~. \label{eq15} \en With this
definition, we have $\Delta=0$ for all choices of $R_1$ and $R_2$
if the free effective string picture of eq.~(\ref{zsmalltot}) is
the correct description of the data. We report in
fig.~\ref{isto80t8} as an example of the results that we obtained
the histogram of $\Delta$ for $\beta=0.75180$, $L=80$  for all the
values of $R_1,R_2\geq 20$. The overall agreement between the data
and the effective string prediction is very good. With the only
exception of the $L=20$ set the data show a rather nice Gaussian
like distribution around the expected $\Delta=0$ value. The
variance of these Gaussians is of the same order of the errors of
$H(R_1,R_2)$ extrapolated from those reported in
tab.~\ref{tabnew2} and \ref{tab1} and no systematic deviation or
asymmetry is visible in the data. These observations suggest that
the (very small) spread around the $\Delta=0$ value is a purely
statistical effect (as a side remark this is also a test of the
fact
 that our statistical errors are reliable).
On the contrary the sample at $L=20$ shows some systematic asymmetry. This is
an indication that for $T\geq T_c/5$ the large distance,
 high temperature effects
discussed in~\cite{chp03} start to give non-negligible contributions within the
precision of our data.

\subsubsection{The zero temperature string tension}
As a byproduct of the previous analysis we may obtain from our
fits a very precise determination of the zero temperature string
tension for three sets of data. They are reported in
tab.~\ref{tabsigma}
 where we have listed the
results of the fits according to eq.~(\ref{eqfitnew2}).

\begin{table}[h]
\begin{center}
\begin{tabular}{|c|c|c|c|c|c|c|}
\hline
$\beta$& $L$& $T/T_c$ &$R_{min}$ & $\chi^2_r$ & $\sigma$ & $c$   \\
\hline
0.65608&20&1/10 &\phantom{0}9& 0.58          &   0.204864(9)\phantom{00}&
  \phantom{0}1.017(11) \\
0.73107&80&1/20          & 10& 1.3\phantom{0}&   0.044023(3)\phantom{00}&
 1.003(4)\\
0.73107&40&1/10          & 10& 2.0\phantom{0}&   0.044019(4)\phantom{00}&
 1.005(5)\\
0.73107&20&1/5\phantom{0}& 10& 0.67          &   0.043985(4)\phantom{00}&
 0.988(6)\\
0.75180&80&1/10          & 22& 1.0\phantom{0}&  0.0105241(15)&
\phantom{0}1.003(10)\\
\hline
\end{tabular}
\end{center}
\caption{\sl Results of the fit according to eq.~(\ref{eqfitnew2}). In the first three
column the values of $\beta$ and $L$ of the data used in the fit and the corresponding
value of $T/T_c$ are summarized.
In the fourth column the minimum value of $R$ included in the
fit is quoted.
In the fifth column we give the reduced $\chi^2$. The last two columns contain
the best
fit results for $\sigma$ and
for the coefficient in front of the effective string correction.}
\label{tabsigma}
\end{table}

 The values of $\sigma$ that we find for $\beta=0.75180$
 is slightly smaller
  than the value we used in our previous
  papers~\cite{cpp02,chp03}, which was
 $\sigma=0.010560(18)$. The effect is very small,
  but nevertheless it is outside
 the quoted error bars (while in the case of $\beta=0.73107$ the error bars are too
 large to detect this effect).  An interesting
 consequence of this (very small) rescaling of $\sigma$ is that it
 could explain the residual (small) systematic
  deviation we found in~\cite{chp03}
 between our data and the Nambu--Goto effective string truncated at the first
 perturbative order (see fig.s~2, 3, 4 and 5 of~\cite{chp03}). Both the sign of
 these deviations
 and the fact that they are proportional to $L$
 are in agreement with this explanation.

\section{Higher order corrections at short distance}
\label{sect3}

With the above results at hand, we are in the position to address
the issue of the higher order correction at short distance in a
reliable and precise way. We shall perform this analysis in two
steps. First we shall study the samples at $\beta=0.73107$ and
$\beta=0.75180$ trying to quantify the higher order terms in the
potential, without referring to any particular effective string
picture. Then, in the second step we shall compare our results
with the expected behaviour of the Nambu--Goto string. In this
second step we shall also include the data at $\beta=0.65608$ in
our analysis.

\subsection{Higher order corrections}
We decided to study the higher order corrections by looking
at the observable (see the def. of eq.~(\ref{eq15}))
$\Delta(R,R-n)$.

This quantity is interesting for several reasons. First of all, as
we mentioned above, it is different from zero only if higher order
corrections are present. Moreover, for $n=1$ we may easily  relate
it with the observable $c(\tilde r)$ introduced in~\cite{lw02}.
The relation is \eq H(R,R-1)=2\frac{c(\tilde r)}{{\tilde r}^3} \en
 (for a definition of $\tilde r$ see~\cite{lw02}). As a side remark,
this means that
 $H(R,R-1)$ (suitably normalized)
 is an estimator of the central charge of the underlying
 conformal field theory (CFT),
(see \cite{chp03} for a discussion of the CFT description of
effective string theory). Thus $\Delta(R,R-1)$
 measures the deviation of the central charge of this CFT
 from the free bosonic value (i.e. $c=1$).

 In order to compare values of this quantity for samples at different values of $\sigma$ and for different
 choices of $n$ it is useful to define the following scale invariant version of $\Delta$
 \eq
 D\left(R+\frac{n+1}{2}\right)\equiv \frac{\Delta(R+n,R)}{n\sigma^{3/2}}
\;\;.
 \en
In building this observable we used the high precision results for
$\sigma$ obtained in the previous section.

We are primarily interested in identifying the exponent of the
correction, thus we fitted $D(R)$ with the ansatz \eq
D(R)=\frac{g_{\alpha+2}}{R^{\alpha+2}} \;\;. \label{el1} \en It is
possible to relate $g_{\alpha+2}$ to a corresponding higher order
correction in the free energy. More precisely, if the free energy
has a correction of the type \eq -\log G(R)\simeq \sigma RL + k(L)
+ \log\left[
\eta\left(i\frac{L}{2R}\right)\right]+\frac{\gamma_\alpha}{\sigma^{\frac{\alpha-1}{2}}}\frac{L}{R^\alpha}
\;\;, \label{eq:19} \en then in $D(R)$ we expect to find a
correction proportional to
 $g_{\alpha+2}/R^{\alpha+2}$ (plus higher order terms
due to the lattice discretization) with $g_{\alpha+2}=\alpha(\alpha+1) \gamma_\alpha$.
Notice that in eq.~(\ref{eq:19}) we have rescaled the coefficient of the higher order correction with a
suitable power of $\sigma$ so as to make
it adimensional. We also implicitly assumed that the higher order term is proportional to
$L$. We shall see below that this assumption agrees with the numerical results.

Looking at the discussion of sect.~\ref{sect1}, it is easy then to
relate the presence of a nonzero $\gamma_\alpha$ term to a precise
contribution in the string action. In particular, if the fit with
$\alpha=1$ gave
 an acceptable (i.e. order unity) reduced $\chi^2$ and a
value $\gamma_1\not=0$, then this would indicate that
the L\"uscher term does not properly
describe the data and that a different coefficient in front of
 the $1/R$ correction to the interquark potential is needed in this short distance
 regime.
If instead the $\alpha=2$ fit gave an acceptable $\chi^2_r$, then
this would indicate that a boundary term might be present in the
data, and we could estimate the boundary parameter $b$ (we shall
discuss the boundary contribution in sect.~\ref{boundary} below).
 Notice that,
according to our normalization choice,
the parameter $b$ introduced by L\"uscher and Weisz in
\cite{lw02} is given by
$b=-\frac{24}{\pi}\frac{\gamma_2}{\sqrt(\sigma)}$.
 We shall discuss this issue in detail below.
Finally, an acceptable $\chi^2_r$ with $\alpha=3$
would indicate that higher order
 terms are present in the effective
 string action. In particular, the
 Nambu--Goto action would precisely give such
 a contribution (see below). The value of $\gamma_3$ extracted from the fit
 could help to identify the form of these higher order terms.
 Clearly, once for a given $\alpha$ we find
 $\gamma_\alpha\not=0$ all the higher powers of $R$ in the fit would
 have nonzero coefficients, regardless of
the presence of a corresponding term
 in the action, simply because of
the lattice artifacts. It is well possible that a mixture of all
these terms is present in the action,
 but unfortunately  our data are not precise enough to allow for more than one
  parameter fits. Thus we shall only be able to identify (if it exists) the leading term  among them. As a
  consequence
  one should look at the results which we shall now discuss
more as a qualitative indication than a
  quantitative estimate of the
higher order terms in the string action. Notwithstanding this, a
few precise
  pieces of information can be obtained from our data.

As a last point let us mention that it is important to properly select the range of values of $R$
to be included in the
fit. According to~\cite{cpp02} and \cite{chp03},
 we must choose $R\geq R_c$ (see the values reported in tab.~\ref{tabnew1}).

We performed the fit for various integer values of $\alpha$; the
results are listed in tab.~\ref{tab2}-\ref{tab2ter} (see also
fig.~\ref{fignew1} and fig.~\ref{fignew2})  for all the data
except those at $L=20$ (which, as we have shown in the previous
section, seem to be affected by high temperature corrections).

\begin{table}
\caption{Results of the fits according to eq.~(\ref{el1})
for the data at $\beta=0.75180$
and $L=80$. In the first column we give the value of $\alpha$, in
the second column the reduced $\chi^2$, in the last two columns the values of
$g_{\alpha+2}$ and $\gamma_\alpha$.}
\label{tab2}
\vskip0.2cm
\begin{tabular}{|c|c|c|r|}
\hline
\multicolumn{1}{|c}{$\alpha$}
&\multicolumn{1}{|c}{$\chi^2_r$}
&\multicolumn{1}{|c}{$g_{\alpha+2}$}
&\multicolumn{1}{|c|}{$\gamma_\alpha$} \\
\hline
$1$ & 2.8 & 0.033(3) &   0.0166(13)\\
$2$ & 1.8 & 0.053(4) &    0.0089(7)\\
$3$ & 1.5 & 0.079(6) &   0.0066(5)\\
$4$ & 1.5 & 0.114(9) &  0.0057(4)\\
\hline
\end{tabular}
\end{table}

\begin{table}
\caption{Same as tab.~\ref{tab2} but for the data at
$\beta=0.73107$ and $L=80$. } \label{tab2bis} \vskip0.2cm
\begin{tabular}{|c|l|c|l|}
\hline
\multicolumn{1}{|c}{$\alpha$}
&\multicolumn{1}{|c}{$\chi^2_r$}
&\multicolumn{1}{|c}{$g_{\alpha+2}$}
&\multicolumn{1}{|c|}{$\gamma_\alpha$} \\
\hline
$1$ & 22.1 & 0.037(1) &   0.0182(4)\\
$2$ & 6.8 & 0.052(1) &    0.0087(2)\\
$3$ & 1.5 & 0.070(1) &   0.0059(1)\\
$4$ & 1.0 & 0.092(2) &  0.0046(1)\\
\hline
\end{tabular}
\end{table}

\begin{table}
\caption{Same as tab.~\ref{tab2} but for the data at
$\beta=0.73107$ and $L=40$.} \label{tab2ter} \vskip0.2cm
\begin{tabular}{|c|c|c|l|}
\hline
\multicolumn{1}{|c}{$\alpha$}
&\multicolumn{1}{|c}{$\chi^2_r$}
&\multicolumn{1}{|c}{$g_{\alpha+2}$}
&\multicolumn{1}{|c|}{$\gamma_\alpha$} \\
\hline
$1$ & 10.4 & 0.036(1) &   0.0180(6)\\
$2$ & 4.1 & 0.052(2) &    0.0087(3)\\
$3$ & 1.8 & 0.070(2) &   0.0058(2)\\
$4$ & 1.4 & 0.092(3) &  0.0046(1)\\
\hline
\end{tabular}
\end{table}
Let us briefly comment on these results.
\begin{description}
\item{1]} The three sets of data show a remarkably similar
pattern. \item{2]} Looking at the $\chi^2$ values it is clear that
our data exclude a $\alpha=1$ correction. This means that the
L\"uscher term we have subtracted perfectly describes the $1/R$
behaviour of the data. \item{3]}
 A boundary term ($\alpha=2$)
 gives  reduced $\chi^2$'s which are higher than the $\alpha>2$ ones in all three cases.
 If we assume
 that the data are described by a boundary correction,
 then the amplitude of such a
 correction turns out to be of the same order of magnitude (and sign) of those that we find in the large
distance regime (see the discussion at the end of
sect.~\ref{boundary}), but the scaling behaviour of the
 large distance results  is not the one expected for a boundary term.
 \item{4]} The fits seem to support a value between 3 and 4 for $\alpha$.
 This result is compatible with a $1/R^3$ ``Nambu--Goto like" correction or, more likely,
 a mixture of a $1/R^3$ plus higher order corrections.
 \item{5]} The two sets of results at $\beta=0.73107$ are compatible within the errors. This means that
 also the higher order corrections in the free energy $F(R,L)$ scale linearly
 with $L$.
 \item{6]} Comparing the sets of data at $\beta=0.73107$ and the one at $\beta=0.75180$ we
 see a good scaling behaviour in $\beta$.
 This is also evident if one looks at fig.~\ref{fignew1} in which the two sets of data are plotted together.

\end{description}

\subsection{Comparison with the Nambu--Goto string}
The most impressive feature of the above analysis is that, although the power that
 we find $1/R^3$ is exactly the one expected according to the Nambu--Goto
action, the value of the
 coefficient $\gamma_3\sim 0.006$ that we find is definitely different
 with respect to the Nambu--Goto prediction
eq.~(\ref{nlo}), namely $\gamma_3=-\frac{\pi^2}{1152}=-0.00856$~,
which is opposite in sign and more or less $4/3$ in magnitude with
respect to our fit's results \footnote{A similar disagreement in
the short distance regime was recently reported
in~\cite{kuti03}.}.

\begin{figure}
\centering
\includegraphics[height=12cm]{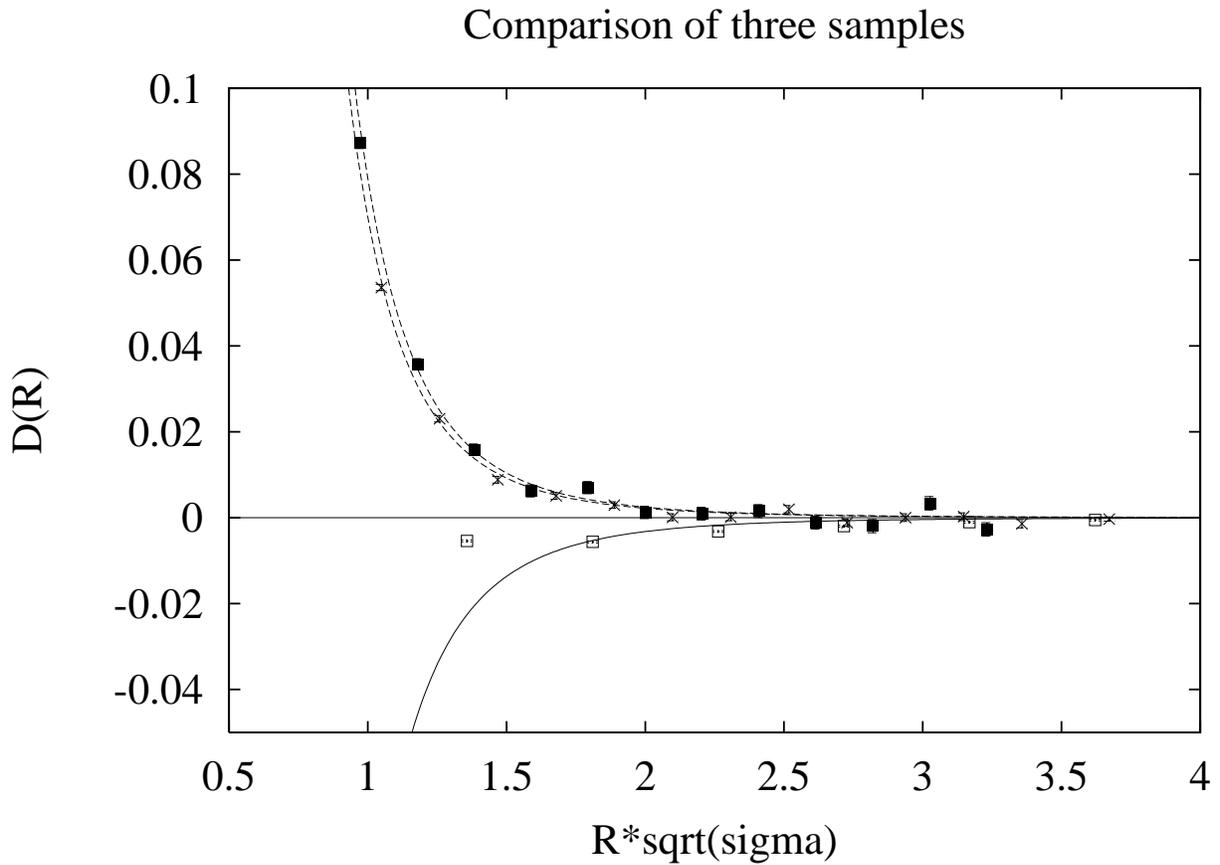}
\caption{Values of $D(R)$ versus $R\sqrt{\sigma}$
for the sample at $\beta=0.65608$ (white squares),
  the one at $\beta=0.75180$ (black squares) and the $L=40$ one at $\beta=0.73107$
(crosses). The continuous line is the Nambu--Goto expectation, the two dashed lines are the best fit results
for $\beta=0.73107$ and $\beta=0.75180$ discussed in the text.}
\label{fignew1}
\end{figure}

\begin{figure}
\centering
\includegraphics[height=12cm]{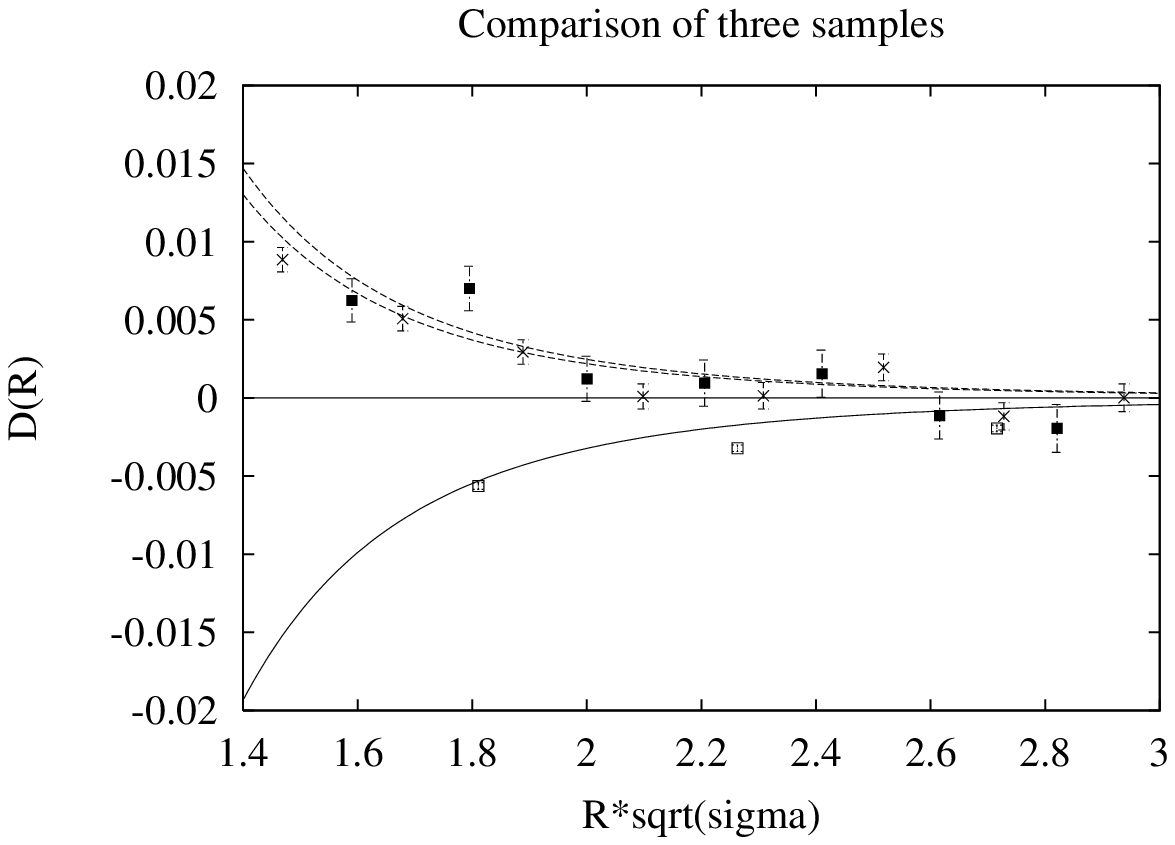}
\caption{Same as fig.~\ref{fignew1} but with a higher resolution.}
\label{fignew2}
\end{figure}

In fig.s~\ref{fignew1} and \ref{fignew2} we plot our data for the
three values of $\beta$ (for $\beta=0.73107$ we chose the sample
with $L=40$). For comparison, we also plot the Nambu--Goto
expected behaviour (solid line) and the two curves obtained using
the two best fit values $\gamma=0.0059$ and $\gamma=0.0066$ for
$\beta=0.73107$ and $\beta=0.75180$ respectively (dashed lines).

Looking at these figures, we see that
 the data for $\beta=0.65608$ show a large scaling violation.
This had to be expected due to the
rather small value of the correlation length $\xi<1$ for this $\beta$.
Note that even the free energy of
a free field theory on a lattice shows corrections of the
order $1/R^3$. As a little exercise, we have computed the free energy
of the free field theory on a square lattice with the appropriate
boundary conditions. We have used the most naive discretisation of the
derivate of the field. We would expect that the artifacts in the Ising model
are at least as large as those in the free field theory on the lattice.
Indeed we find that the corrections in the free field theory have the same
sign as those of the Ising gauge model at $\beta=0.65608$,
but a smaller amplitude.

On the other hand, the fact that the deviations between
$\beta=0.73107$ and $\beta=0.75180$ are much smaller than those
with $\beta=0.65608$, makes us confident that for $\beta>0.73$
lattice artifacts have only a minor impact on the higher order
corrections and hence we can extract continuum physics results.

A possible interpretation of this scenario,
keeping also into account the results of our previous
paper~\cite{chp03}
is that,
while the Nambu--Goto effective string provides a  
correct description
of finite temperature effects
of the interquark potential at large distances, while it fails 
to describe corrections at small distance $R$.
Along with Dirichlet boundary conditions, 
some irrelevant operator becomes important and drastically
modifies the shape of the interquark potential
(or better: the shape of its deviation from the free
string behaviour).

The simplest candidate for such a non-universal behaviour is a
boundary term of the type discussed in the next section. A
positive and large enough value of $b$ could perhaps justify the
observed behaviour of $D(R)$. However as we shall see in the next
section the large distance behaviour of the interquark potential
in the Ising case seems to suggest that $b\sim 0$. Thus it is
possible that some other operator, maybe an extrinsic curvature
contribution~\cite{ext,oy86} is actually responsible for the
observed behaviour. In order to better understand this point we
shall devote the following two sections to a detailed discussion
of the boundary term.

An alternative scenario is %on the other hand, 
that the short distance breakdown of the Nambu--Goto string description
has to be explained by effects beyond an effective string theory;
for a discussion we refer the reader to ref. \cite{kuti03}.

\section{The boundary term}
\label{boundary} As we have seen in the previous section, in order
to understand the short distance behaviour of the effective string
potential it would be important to have an independent estimate of
the boundary parameter $b$. This can be achieved by studying the
large distance data. However, in order to perform this analysis,
we must know the functional form of the boundary correction in the
large distance regime. This section is devoted to such
calculation, while in the next section we shall compare this
result with the large distance data of ref.~\cite{chp03}.

\subsection{The functional form of the boundary correction}

As mentioned in the introduction, the most general effective
string action in presence of boundaries requires the inclusion in
the action of terms that are localized at the boundary. The
simplest possible term of this type is~\cite{lw02}

\eq \label{bterm} \mathcal{A}_b=\frac{b}{4} \int_0^L \!\!\! dt
\left[  \left( \frac{\partial h}{\partial z} \right)_{z=0}^2 +
\left( \frac{\partial h}{\partial z} \right)_{z=R}^2 \right] \;\;,
\en where $b$ is a parameter with dimensions $[\mbox{length}]$,
$h$ denotes the transverse displacement of the string (see the
appendix for a detailed discussion) and the $\frac{1}{4}$ factor
has been added to agree with the conventions of \cite{lw02}.

This additional term can be treated in the framework of the
zeta-function regularization in a way similar to the Nambu--Goto
one. We report in the appendix a detailed account of this
calculation and only list here the final result, which turns out
to be rather simple.

At first perturbative order in $b$, the regularization of the free
string action plus the boundary term gives the same result as the
pure free string action, namely the Dedekind function discussed
above, provided one replaces the interquark distance $R$ with \eq
R \to R^*=\frac{R}{(1+2\frac{b}{R})^{\frac12}} \en (remember that
$b$ has the dimension of a length).

Thus, if we denote the effective string contribution to the interquark potential
at this order (with an obvious choice of notation)
as  $F_q^1(R,L)+F_q^b(R,L)$,
 we find (we shall keep from now on $d=3$ to simplify the
 equations)
 \eq
 F_q^1(R,L)+F_q^b(R,L)=\log \eta \left( i \frac{L}{2R^*} \right) \;\;.
 \en
Expanding this result in the short distance regime (i.e. for $2R^*<L$),
and neglecting the exponentially decreasing corrections we find
 \eq
 F_q^1(R,L)+F_q^b(R,L)=-\frac{\pi L}{24
 R}\left(1+2\frac{b}{R}\right)^{\frac12} \;\;,
 \en
which becomes at the first order in $b$
 \eq
 F_q^1(R,L)+F_q^b(R,L)=-\frac{\pi L}{24 R}\left(1+\frac{b}{R}\right) \;\;,
 \en
which is exactly L\"uscher and Weisz's result~\cite{lw02}.

\subsection{Large distance behaviour of the boundary correction}
The major advantage of having the complete functional form of the boundary
correction is that we can now look at its large distance behaviour.
By using the modular transformation of the Dedekind function
we find for $2R^*>L$

\eq
 F_q^1(R,L)+F_q^b(R,L)=\left[-\frac{\pi R^*}{6 L}+
\frac{1}{2} \log\frac{2R^*}{L}
+\sum_{n=1}^\infty \log (1-e^{-4\pi nR^*/L})\right]~~~.
\label{eq2}
\en
Neglecting again the exponentially decreasing terms, and
keeping only the first
order in $b$ we end up with

\eq
 F_q^1(R,L)+F_q^b(R,L)=\left[-\frac{\pi R}{6
 L}\left(1-\frac{b}{R}\right)+
\frac{1}{2} \log\frac{2R}{L} - \frac14\log\left(1+2\frac{b}{R}\right)\right]
\;\;,
\label{eq3}
\en

which means (remember again that we only keep the terms
proportional to $b$ in the expansion)

\eq
 F_q^1(R,L)+F_q^b(R,L)=-\frac{\pi R}{6 L}+\frac{\pi b}{6L}+
\frac{1}{2} \log\frac{2R}{L} - \frac{b}{2R}
\label{eq4}
\en

from which we obtain

\eq
F_q^b(R,L)= +\frac{\pi b}{6L} - \frac{b}{2R} ~~.
\en

Thus we see that the effect of the boundary term is to renormalize the constant
term $k(L)$ (which however is not relevant in the ratios
evaluated in Monte Carlo simulations) and to add a further $1/R$ correction.

 A few observations are in order at this point.
 \begin{itemize}
 \item
 This result is rather satisfactory from a physical point of view: the effect of
 a boundary term can only decrease as a function of $R$ (i.e. when the two
 boundaries are far apart), and this is indeed the case in the large $R$
 regime, too. This is in sharp contrast with the behaviour of the
 string fluctuations, whose dominant term in the large distance
 regime is always (i.e. at any order in the $1/{\sigma RL}$ expansion) a
 linearly rising correction.
 \item
 Although both in the short distance and in the large distance regimes
 the correction induced by the boundary term
has the same sign,
 nevertheless there is a well defined change of behaviour: from
  $-\frac{b\pi L}{24 R^2}$ to $-\frac{b}{2R}$. This feature may simplify the
  identification of such a term  in numerical simulations.
  Notice in particular the lack of $L$
  dependence in the large $R$ regime.
  \end{itemize}

\section{Comparison with the Ising gauge model in the large distance regime}
\label{sect2} The aim of this section is to compare the prediction
of eq.~(\ref{eq4}) with the results of the simulations reported
in~\cite{chp03} for the large distance regime of the interquark
potential in the 3d gauge Ising model. In particular, we shall
concentrate on the set of simulations performed at
$\beta=0.75180$, which were the most precise ones.  The
simulations were performed on lattices with space-like size
$N_s=128$ and ``time'' sizes $L=10,12,16,24$, corresponding to
 $\frac45T_c,~\frac23T_c,~\frac12T_c$ and $\frac13T_c$,
respectively.
For each temperature, we studied the correlators for various values of the
interquark distance $R$, ranging from $R=8$ to $R=48$. The data are reported
in~\cite{chp03}.

The most effective observable to identify the possible presence of
a boundary correction is the following combination \eqa Q_2(R)
&\equiv&  -\log\frac{G(R+1)}{G(R)} - F_q^1(R+1) + F_q^1(R)
\nonumber \\
       &=&  -\log\frac{G(R+1)}{G(R)} +\log\left[\eta\left(i\frac{L}{2R}\right)\right]
-\log\left[\eta\left(i\frac{L}{2(R+1)}\right)\right] \;\;.
\label{defq2}
\ena

In order to understand the results of the fit it may be useful to
extract from eq.s~(\ref{bos}) and (\ref{nlo}) the expected
behaviour in the large distance regime for this quantity (which as
we mentioned several times is  drastically different from the
short distance one discussed in sect.s~\ref{sectnew} and
\ref{sect3}). Discarding exponentially decreasing terms, we find
\eq F_q^1(R+1,L)-F_q^1(R,L)=-\frac{\pi}{6L}+\frac12
\log\left(\frac{R+1}{R}\right) \label{eqlr1} \en and (see
eq.s~(2.37) of ref. \cite{chp03}) \eq
F_q^2(R+1,L)-F_q^2(R,L)=-\frac{\pi^2}{72\sigma L^3}
-\frac{1}{8\sigma L}\frac{1}{R(R+1)} \;\;. \label{eqlr2}\en These
corrections should be compared with what is expected from the
boundary term, namely \eq
F_q^b(R+1,L)-F_q^b(R,L)=\frac{b}{2R(R+1)} \;\;. \label{eqlrb}\en
The boundary correction to $Q_2(R)$ has the same $R$ dependence as
the
 Nambu--Goto term, but a {\bf
different $L$ dependence} (this is an obvious consequence of the fact that $b$ is a
dimensional parameter): that makes it very easy to disentangle between the
two corrections. In particular, fitting them with the data
of \cite{chp03}, we
find that $b$ is compatible with zero, i.e. that no boundary correction seems to
be present in the 3d Ising gauge model.

More precisely, we fitted $Q_2(R)$, for each of the four values of
$L$, with the following formula \eq Q_2(R)=a_0+b_0\frac{1}{R(R+1)}
\;\;, \label{el0} \en where $a_0$ is related to the string
tension, while $b_0$ encodes the large distance $1/R$ correction
to the potential we are interested in. With this choice of
normalization, if the $1/R$ correction is completely due to the
boundary term, then $b=2b_0$. Since for this value of $\beta$ we
have $R_c\sim 12$, we only fitted the data for $R\geq 16$, for a
total of 6 points.  The results of the fits are reported in
tab.~\ref{tab0}.

\begin{table}[h]
\caption{\sl Results of the fits to eq.~(\ref{el0}).
In the first column the value of $L$, in
the second column the reduced $\chi^2$ is given,
in the last two columns the values of
$b_0$ and the corresponding expectation --- see eq.~(\ref{eqlr2})
--- according to the Nambu--Goto
string.\label{tab0}} \vskip0.2cm
\begin{tabular}{|c|c|c|c|}
\hline
$L$ &$\chi^2_r$ & $b_0$ & $1/8\sigma L$   \\
\hline
$10$ & 2.72 & -1.31(7) &  -1.184\\
$12$ & 2.33 & -0.97(6) &  -0.986\\
$16$ & 2.30 & -0.62(5) &  -0.740\\
$24$ & 2.54 & -0.30(4) &  -0.493\\
\hline
\end{tabular}
\end{table}

Few observation can be made on these results
\begin{description}
\item{a]} The magnitude of the reduced $\chi^2$ (which is
remarkably constant as $L$ changes) indicates that in the large
distance regime the Nambu--Goto string (truncated at the first
perturbative order) is a reasonable approximation, but does not
fully describe the data. \item{b]} There is a clear $L$ dependence
in the best fit values of $b_0$. These values qualitatively agree
with the Nambu--Goto expectation, but they seem to show an even
steeper dependence on $L$. Apparently little room is left for a
boundary correction term. A naive extrapolation to $L\to\infty$
suggests a very small value of $b$, most probably compatible with
zero. \item{c]} It is interesting to compare what we find here
with the fits discussed in sect.~\ref{sect3} (see
tab.s~\ref{tab2}, \ref{tab2bis} and
 \ref{tab2ter})
 we find for this value of $\beta$:
 $b_0\equiv b/2=-\frac{12}{\pi}\frac{\gamma_2}{\sqrt{\sigma}}\sim-0.35$
which is of the same order of magnitude and has the same sign of
the values reported in tab.~\ref{tab0}.
\end{description}

\section{Concluding remarks}
\label{sect4}
Let us summarize the main results of our analysis.
\begin{description}

\item{a]} We obtained an explicit form for the boundary
correction, at first perturbative order in $b$, which is valid for
any value of $R$ and $L$. The importance of this result is that in
the large $R$ regime, by virtue of the peculiar form of the
boundary correction, i.e. the fact that it is $L$-independent, it
is much simpler to measure the value of $b$.

\item{b]} As a first application of this result,
we evaluated $b$ in the 3d
gauge Ising model, using the large distance data published in~\cite{chp03}. It
turns out that $b$ is very small and compatible with zero.

 \item{c]} In order to test the effective string
 picture
in the short distance regime as well, we performed a new set of
simulations
for the 3d gauge Ising
model. A careful analysis of the data in the range of $T$ and $R$ values 
in which higher
order terms can be neglected shows a perfect agreement with the prediction 
of the free bosonic string.
However it turns out that in order to describe the data one must
keep into account {\bf the whole functional form of the correction}. 
Cutting it to the
 L\"uscher term only may induce a relevant mismatch with the data.

 \item{d]}
 Finally we studied the higher order corrections for small values of $R$.
   In this regime the Nambu--Goto correction, truncated at the first
 perturbative order, should behave as $1/R^3$. Assuming $b=0$ and
 fitting the data with this correction (the $\alpha=3$ fit
 in the sect.~\ref{sect3}) we found an acceptable $\chi^2_r$,
 but the wrong coefficient.

 \end{description}

 This is indeed 
rather a puzzling
result since in the large distance, high 
temperature regime we found 
instead 
quite a good agreement with
the Nambu--Goto prediction (see \cite{chp03}). In
principle, a description holding at short distances could even be a 
picture different from an effective string \cite{kuti03}. 
However, it is also possible that the short distance breakdown
of the Nambu--Goto string scenario 
could be explained within the framework of an effective string theory.
In particular,  following the comments that we made
at the end of sect.~\ref{sect3}, we see  two  
possible explanations for this behaviour.
\begin{description}
 \item{1]}
The  Nambu--Goto picture could be the wrong assumption. 
In this framework the good
agreement at large distance is only a coincidence, and one should look for some
other, more exotic,  effective string action to simultaneously describe the
short distance as well as the finite temperature
behaviour of the potential. Notice however that none of the generalized
strings discussed in \cite{df83} satisfies these constraints, 
thus some drastically different
model would be needed to follow this line.

\item{2]} 
The 
short distance deviation with respect to the Nambu--Goto
prediction  
might be due to 
an irrelevant operator which may possibly have a string-like
description. Two natural options are:
\begin{itemize}
\item
 a higher order boundary term which decreases
with a higher power of $R$ and could thus be compatible both with the behaviour that we extract from the short
distance fits (which suggests a power higher than $1/R^2$) and with the large distance regime.
\item a term proportional to the extrinsic curvature of the string~\cite{ext,oy86}. It is possible that such
a term gives significantly different contributions in presence of Dirichlet 
and periodic boundary conditions.
\end{itemize}

In this framework it could be useful to compare the short distance
behaviour of different models to see if some common feature emerges. The nice
feature of this last option is
that it does not require a large value of the $b$ parameter and thus it
does agree with all our numerical
results.

\end{description}

\begin{figure}
\centering
\includegraphics[height=12cm]{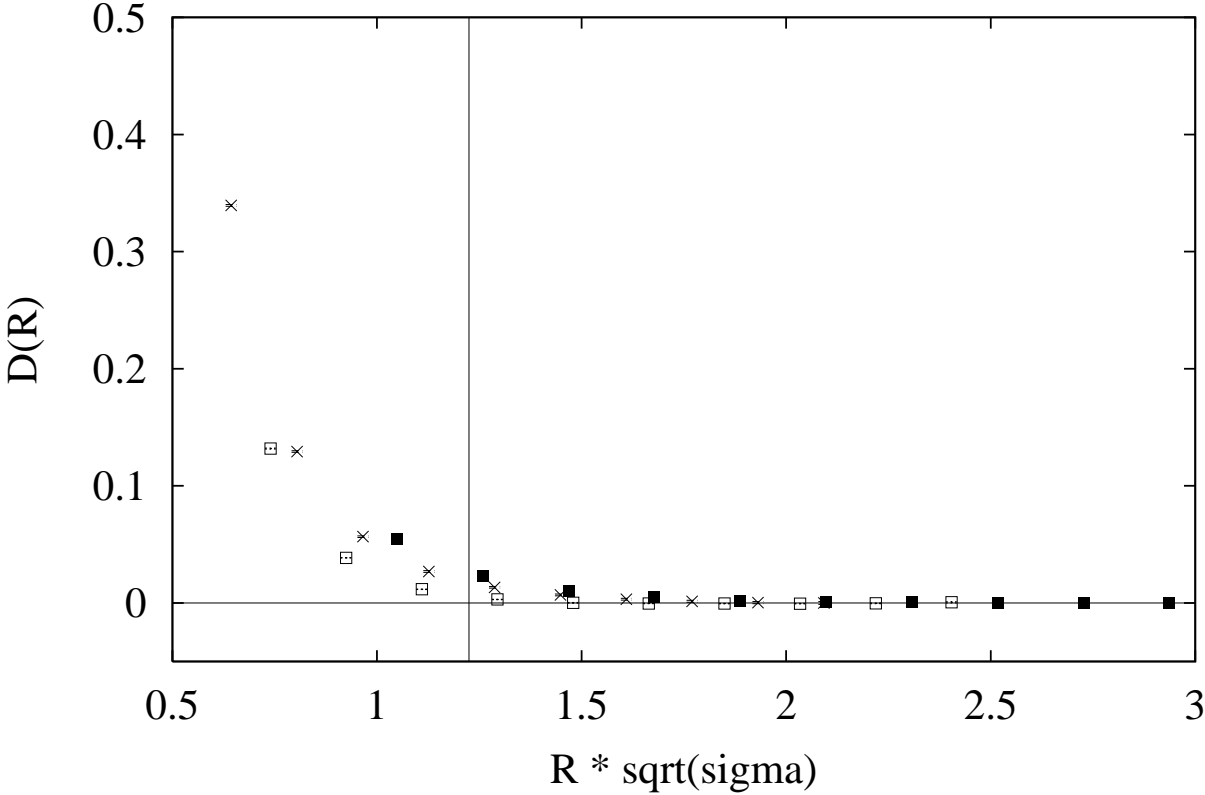}
\caption{Values of $D(R)$ versus $R\sqrt{\sigma}$
for the d=3 SU(3) gauge theory (white squares),
  the Ising
gauge model (black squares) and the 3d SU(2) gauge theory (crosses).
The vertical line corresponds to $R=R_c$.}
\label{fig3}
\end{figure}

\begin{figure}
\centering
\includegraphics[height=12cm]{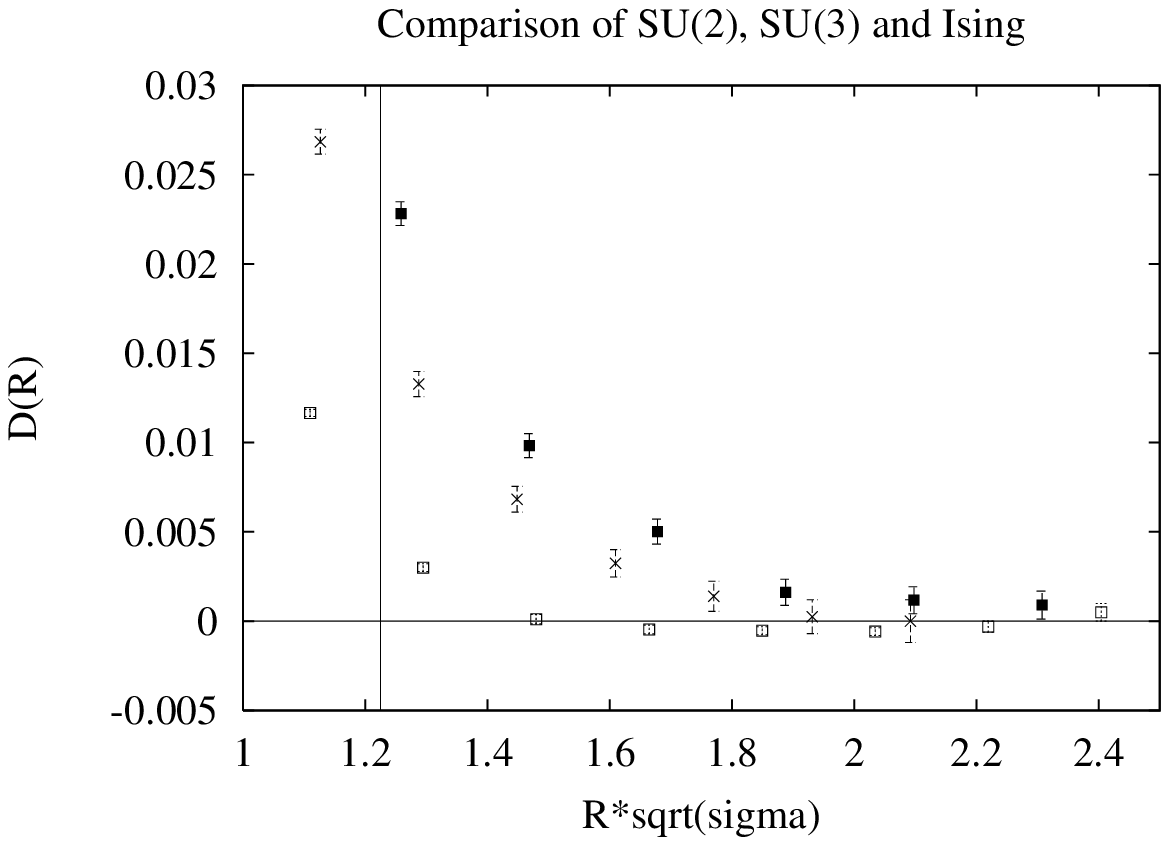}
\caption{Same as fig.~\ref{fig3} but with a higher resolution.}
\label{fig3zoom}
\end{figure}

Following the suggestion of the last point above we tried to test
if some common behaviour is shared by different gauge theories. To
this end, we performed the same analysis discussed in
sect.~\ref{sect3} with the data on the 3d SU(3) gauge model
published in~\cite{lw02}. In particular,
 we concentrated on the sample at $\beta=20$, see the
 data in tab.~3 of ref.~\cite{lw02},
which correspond to a value $\sigma\sim 0.0342$. In agreement with
 the results of~\cite{lw02}, we found that the data are well
described by  the free string correction only, with apparently no need of higher
order corrections. The most impressive signature of this behaviour
 is given by the observable $D(R)$ introduced in sect.~\ref{sect3}
  which is different from zero only if higher order corrections are
present. It turns out that for SU(3), $D(R)$ is  almost compatible
with zero within the errors, for all of the values
$\sqrt{\sigma}R>1.3 $. To complete the comparison we also studied
a set of data obtained in the 3d SU(2) model taken
from~\cite{cpr03}. These data are characterized by a slightly
smaller string tension $\sigma\sim 0.0263$ and a temperature
$T/T_c\sim 1/7$ (see~\cite{cpr03} for details). The value of
$D(R)$ for these two sets of data, together with the  $L=80$,
$\beta=0.73107$ sample of the Ising model are reported in
fig.s~\ref{fig3} and \ref{fig3zoom}. Notice that the SU(3) and
SU(2) samples have values of $\sigma$ which are smaller than the
Ising one. Looking at these figures one  sees that the  SU(2) data
lie somehow in between the Ising and the SU(3) ones. This is also
confirmed by the $\gamma_3$ value extracted from the SU(2) data:
$\gamma_3\sim 0.004$~\cite{cpr03} which is smaller (and outside
the error bars) than the one that we obtained in the present paper
for the Ising case.

A similar comparison between the 3D SU(2) and 
$\Z_2$ gauge models
for the first 
excited string level can be found in ref. \cite{kuti03}. 
In this case, the SU(2) and 
$\Z_2$ data seem to be compatible.

Further studies are needed to better characterize these
behaviours, to understand if they may be related to some feature
of the gauge group (say, for instance, to the center of the group
as it is suggested by the analogy between Ising and SU(2)
behaviours) and to see if they support any of the scenarios
proposed above.

\vskip1.0cm {\bf
Acknowledgements.} This work was partially supported by the
European Commission TMR programme HPRN-CT-2002-00325 (EUCLID). M.~P.
acknowledges support received from Enterprise Ireland under the 
Basic Research Programme. 
M.~H. thanks PPARC for support under grant PPA/G/O/2002/00468.
%%%%%%%%   FIRST REVISED VERSION
%%%%%%%%   MARCO, MAY 12TH
%%%%%%%%
%%%%%%%%
\newpage
\appendix{}
\label{app}
\vskip 0.5cm
\section{{{Evaluation of the functional form of the boundary correction.}}}
%%%%%%%%%%%%%%%%%%%%%%%%%%%%%%%%%
\renewcommand{\theequation}{A.\arabic{equation}}
\setcounter{equation}{0}

Let us begin by considering the case when we assume
that the dynamics of the effective
string world sheet is described by a pure Nambu--Goto
action, neglecting a possible boundary term: under this
hypothesis, the action is given by the string world
sheet surface area (measured in units of the inverse
string tension), and the expectation value for the Polyakov
loop correlator can be expressed as
the partition function:
\eq
\label{partitionfunction}
Z= \int \left[ \mathcal{D} h \right] e^{-\sigma \mathcal{A}_0}
\qe
The Nambu--Goto action
$\mathcal{A}_0$ can be written in the following form:
\eq
\label{nambugoto}
\mathcal{A}_0 = \int_0^R \!\!\! dz \int_0^L \!\!\! dt
\sqrt{\det g_{\alpha \beta}}
\qe
where $g_{\alpha \beta}$ is defined as:
\eq
\label{g}
g_{\alpha \beta} = \partial_\alpha X^\mu\partial_\beta X_\mu
\en
Let us focus on the case of three dimensions:
we can parametrize the $X$ field in the
following way:
\eqa
\label{parametrizzazione}
X_1&=&z; \\ \nonumber
X_2&=&t; \\ \nonumber
X_3&=&h \;\;=\;\; h\left(z,t\right); \\
\ena
and the $h$ field is associated to the string world sheet
surface transverse displacement with respect to a minimal area
plane. Thus we have:
\eq
\label{nambugoto3d}
\mathcal{A}_0 = \int_0^R \!\!\! dz \int_0^L \!\!\! dt
\sqrt{1 + \left( \frac{\partial h}{\partial z} \right)^2
+\left( \frac{\partial h}{\partial t} \right)^2}
\en
A Taylor expansion of the square root
appearing in Eq. (\ref{nambugoto3d})
gives --- to first non-trivial order:
\eq
\label{aquadratica}
\mathcal{A}_0 \simeq \int_0^R \!\!\! dz \int_0^L \!\!\! dt
\left\{
1+\frac{1}{2} \left[ \left( \frac{\partial h}{\partial z} \right)^2
+\left( \frac{\partial h}{\partial t} \right)^2 \right]
\right\}
\en
and the partition function can be approximated as:
\eq
\label{z0+1}
Z_{0+1}= e^{-\sigma RL} \int \left[ \mathcal{D} h \right]
e^{-\frac{\sigma}{2} \int_0^R \!\!\! dz \int_0^L \!\!\! dt
\left[ \left( \frac{\partial h}{\partial z} \right)^2
+\left( \frac{\partial h}{\partial t} \right)^2 \right]}
\en
On the other hand, if we allow for the existence of a possible ``boundary
term'', the effective action will take the form:
\eq
\label{zstar}
Z^\star = \int \left[ \mathcal{D} h \right] e^{-\sigma \mathcal{A}}
\en
with:
\eq
\label{a}
\mathcal{A}=\mathcal{A}_0 +\frac{b}{4} \int_0^L \!\!\! dt
\left[  \left( \frac{\partial h}{\partial z} \right)_{z=0}^2
+  \left( \frac{\partial h}{\partial z} \right)_{z=R}^2
\right]
\en
where $b$ is a parameter with dimensions $[\mbox{length}]$,
while 
the $\frac{1}{4}$ factor is just a conventional 
choice (see also \cite{lw02}).\\
Let us notice that, in the case of the system we are
presently interested in, namely: a finite
temperature confining gauge theory\footnote{The string world sheet
associated to the expectation value of the Polyakov
loop correlator in a finite temperature gauge theory
is characterized by fixed boundary conditions in the
$z$ direction, and periodic boundary conditions
in the compactified, $t$ direction.}, the $z$-dependence of 
the eigenfunctions appropriate to the case of the pure
Nambu--Goto action can be factored out as:
\eq
\label{zdependence}
\sin \left( \frac{n\pi z}{R} \right)  \;\;\; , \;\;\;n \in \mathbf{N}_0
\en
thus:
\eq
\label{autovalori}
\frac{\partial h}{\partial z}= \frac{n\pi}{R} \cos \left(
\frac{n\pi z}{R} \right) \dots
\en
where we omitted the eigenfunction term
expressing the $t$-dependence; thus we have:
\eqa
\label{autovaloriin0er}
\left( \frac{\partial h}{\partial z} \right)_{z=0}^2 \!\!\!\!\!\!&=&\!\!
\left( \frac{\partial h}{\partial z} \right)_{z=R}^2 =
\frac{n^2 \pi^2}{R^2} \left. \cos^2 \left( \frac{n\pi z}{R}
\right) \right|_{z=0} \dots =
\frac{n^2 \pi^2}{R^2} \left. \cos^2 \left( \frac{n\pi z}{R}
\right) \right|_{z=R} \dots \nonumber \\
\!\!\!&=&\!\! \frac{n^2 \pi^2}{R^2} \dots
\ena
As it is concerned with the momentum in the compactified
direction, the eigenvalues are:
\eq
\label{autovalorit}
\frac{4\pi^2 m^2}{L^2} \;\;\; , \;\;\;m \in \mathbf{Z}
\en
If we define:
\eq
\label{rhotau}
\left\{
\begin{array}{c}
\rho = R \\
\tau = \frac{L}{2}
\end{array}
\right.
\en
the eigenvalues of the $(-\Delta)$ operator can be written as:
\eq
\label{autovalori-lapl}
\pi^2 \left( \frac{n^2}{\rho^2} + \frac{m^2}{\tau^2} \right)
\qe
Taking into account the boundary term too, the
complete action $\mathcal{A}$ reads, to first non-trivial order:
\eqa
\label{acompletaquadratica}
\mathcal{A} \simeq & & \!\!\!\!\!\!\!\! \int_0^R \!\!\! dz \int_0^L \!\!\! dt
\left\{
1+\frac{1}{2} \left[ \left( \frac{\partial h}{\partial z} \right)^2
+\left( \frac{\partial h}{\partial t} \right)^2 \right]
\right\} \nonumber \\ 
& &+ \frac{b}{4} \int_0^L \!\!\! dt
\left[  \left( \frac{\partial h}{\partial z} \right)_{z=0}^2 \!\!\!\!
+  \left( \frac{\partial h}{\partial z} \right)_{z=R}^2
\right]
\ena
Now, we propose to work out the correction to the 
$V(R)$ interquark potential to first order in $b$ 
in a perturbative expansion; to do this, we have to 
evaluate:
\eq
\label{z01}
Z_{0+1}^\star = \left[  \int \left[ \mathcal{D} h \right]
e^{-\sigma \mathcal{A}} \right]_{\mbox{\tiny{up to first order}}}
\en
The Taylor expansion of the exponent gives:
\eqa\label{sigmaa}
\sigma \mathcal{A} \simeq & &\!\!\!\!\!\!\!\! \sigma RL + \frac{\sigma}{2}
\int_0^R \!\!\! dz \int_0^L \!\!\! dt 
\left[ \left( \frac{\partial h}{\partial z} \right)^2 \!\!\!
+\left( \frac{\partial h}{\partial t} \right)^2 \right] \nonumber \\ 
& & + \frac{\sigma b}{4} \int_0^L \!\!\! dt
\left[  \left( \frac{\partial h}{\partial z} \right)_{z=0}^2 \!\!\!\!
+  \left( \frac{\partial h}{\partial z} \right)_{z=R}^2
\right]
\ena
We can think about this formula as the starting point
in a perturbative analysis in terms of the $b$
parameter, namely we
can consider the boundary term as a perturbation
of the pure Nambu--Goto action Eq. (\ref{nambugoto}).\\
First of all, let us notice that:
\begin{enumerate}
\item by virtue of the fact that the eigenvalues
of the unperturbed problem are known --- see
Eq. (\ref{autovalori-lapl}) --- the double integral appearing in 
Eq. (\ref{sigmaa}) can be written in a simpler way:
\eqa
\label{duezampe}
\int_0^R \!\!\! dz \int_0^L \!\!\! dt 
\left[ \left( \frac{\partial h}{\partial z} \right)^2 \!\!\!
+\left( \frac{\partial h}{\partial t} \right)^2 \right] =
\nonumber \\
= \pi^2 \left( \frac{n^2}{\rho^2} + \frac{m^2}{\tau^2} \right)
\cdot \int_0^R \!\!\! dz \int_0^L \!\!\! dt 
\left( \dots \right)_{[z,t]^2}
\ena
where the notation $\left( \dots \right)_{[z,t]^2}$ 
represents (the square modulus of) the complete,
unperturbed eigenfunction.
\item On the other hand, Eq. (\ref{autovaloriin0er}) shows 
that the two addends 
appearing in the integral associated to the
boundary term in Eq. (\ref{sigmaa}) are equal, and 
they give rise to a contribution which can be 
written as:
\eqa\label{unazampa}
\int_0^L \!\!\! dt
\left[  \left( \frac{\partial h}{\partial z} \right)_{z=0}^2 \!\!\!\!
+  \left( \frac{\partial h}{\partial z} \right)_{z=R}^2
\right] = \nonumber \\
= 2 \frac{n^2 \pi^2}{R^2} \cdot \int_0^L \!\!\! dt
\left( \dots \right)_{[t]^2}
\ena
In this case, the integrand $\left( \dots \right)_{[t]^2}$
is just (the square modulus of) the $t$-dependent
part of the
eigenfunction\footnote{The
$z$-dependence does not appear, because we already 
evaluated its value in $z=0$ and in $z=R$.}.
\end{enumerate}
Thus, in the ``pure Nambu--Goto'' case,
$\sigma \mathcal{A}_0$ involves terms like:
\eq\label{terminipurenambugoto}
\frac{\sigma}{2} \pi^2 \left( \frac{n^2}{\rho^2} + \frac{m^2}{\tau^2} \right)
\cdot \int_0^R \!\!\! dz \int_0^L \!\!\! dt 
\left( \dots \right)_{[z,t]^2}
\en
but when the boundary term is included this
expression has to be replaced by:
\eq\label{terminicompleta}
\frac{\sigma}{2} \pi^2 \left( \frac{n^2}{\rho^2} + \frac{m^2}{\tau^2} \right)
\cdot \int_0^R \!\!\! dz \int_0^L \!\!\! dt 
\left( \dots \right)_{[z,t]^2} + \frac{\sigma b}{2}
\frac{n^2 \pi^2}{R^2} \cdot \int_0^L \!\!\! dt
\left( \dots \right)_{[t]^2}
\en
The integrals appearing in Eq. (\ref{terminicompleta})
give rise to results\footnote{The results appearing in
Eq. (\ref{nonnormalizzata2}) and Eq. (\ref{nonnormalizzata1})
refer to an unnormalized
eigenfunction, but this is not relevant to our present
calculation, since the only difference in the normalized
eigenfunction case is that both quantities are 
multiplied by a common factor, and their ratio is
unchanged.} in the form:
\eqa
\label{nonnormalizzata2} 
\int_0^R \!\!\! dz \int_0^L \!\!\! dt 
\left( \dots \right)_{[z,t]^2} &=& C_2 \cdot RL \\
\label{nonnormalizzata1}
\int_0^L \!\!\! dt
\left( \dots \right)_{[t]^2} &=& C_1 \cdot L
\ena
where $C_2$ and $C_1$ are just pure numbers\footnote{Strictly 
speaking, this is true only if we do not consider
the usual normalization factor; however, even in the
more general case, the ratio between $C_2$ and $C_1$
will be the same as we find here.}; thus we can rewrite 
Eq. (\ref{terminicompleta}) as:
\eqa
\label{conglobare}
& & \sigma \frac{\pi^2}{2} \left( \frac{n^2}{\rho^2} + \frac{m^2}{\tau^2} \right)
\cdot C_2 RL + \frac{\sigma b}{2}\pi^2
\frac{n^2}{R^2} \cdot C_1 L = \nonumber \\
& & = \sigma RL \frac{\pi^2}{2}C_2 \cdot 
\left[ \left( \frac{n^2}{\rho^2} + \frac{m^2}{\tau^2} \right) +
\frac{b}{R} \cdot \frac{C_1}{C_2} \cdot \frac{n^2}{\rho^2} 
\right] 
\ena
Defining:
\eq\label{k}
k=\frac{C_1}{C_2}
\en
then Eq. (\ref{conglobare}) can be rewritten as:
\eqa
\label{riscritturacompleta}
& & \sigma \frac{\pi^2}{2} \cdot C_2 RL \left[
\frac{m^2}{\tau^2} + \frac{n^2}{R^2} \left(
1 + k \frac{b}{R}
\right) \right] = \nonumber \\
& & = \sigma \frac{\pi^2}{2} \cdot C_2 RL \left\{
\frac{m^2}{\tau^2} + \frac{n^2}{\left[
R \left( 1 + k \frac{b}{R} \right)^{-\frac{1}{2}}
\right]} \right\}
\ena
By defining:
\eq\label{rstar}
R^\star = R \left( 1 + k \frac{b}{R} \right)^{-\frac{1}{2}}
\en
Eq. (\ref{riscritturacompleta}) takes the form:
\eq
\label{esattamenteuguale}
\sigma \frac{\pi^2}{2} \cdot C_2 RL \left(
\frac{m^2}{\tau^2} + \frac{n^2}{{R^\star}^2}
\right)
\en
In the pure Nambu--Goto theory (with no boundary term),
the corresponding term:
\eq
\label{corrispondente}
\sigma \frac{\pi^2}{2} \cdot C_2 RL \left(
\frac{m^2}{\tau^2} + \frac{n^2}{R^2}
\right)
\en
is associated to the $Z_1$ term in the partition function
factorization formula Eq. (\ref{z0+1}); in three dimensions,
$Z_1$ turns out to be equal to:
\eq\label{z1}
Z_1 = \left[ \eta \left( i \frac{\tau}{\rho} \right) \right]^{-1}
= \left[ \eta \left( i \frac{L}{2R} \right) \right]^{-1}
\en
and its (additive) contribution to the interquark
potential is:
\eq\label{v1}
V_1(R) = - \frac{1}{L} \log Z_1 = \frac{1}{L} \log \eta \left( i \frac{L}{2R} \right)
\en
Since Dedekind's function is defined as:
\eq
\label{eta}
\eta \left( i \frac{\tau}{\rho} \right) = q^\frac{1}{24}
\prod_{n=1}^{+\infty} (1-q^n) \;\;\;\;\; , \;\;\;\;\;
q=e^{-2\pi \frac{\tau}{\rho}}
\en
In the regime $\frac{L}{2R} \gg 1$, we can approximate:
\eqa \label{luescherterm}
V_1(R) \!\!\!\!\!\!\!\! & & = \frac{1}{L} \left[
-\frac{\pi L}{24 R} + \sum_{n=1}^{+\infty} \log
\left( 1 - e^{-\frac{\pi L}{R}n} \right)
\right] \simeq \nonumber \\
& & \simeq - \frac{\pi}{24 R}
\ena
which is nothing but the celebrated
``L\"uscher term'' \cite{lsw}.\\
When the boundary term is included in the action,
we get a similar result, provided we replace:
\eq\label{sostituzione}
R \longrightarrow R^\star = R \left( 1 + k \frac{b}{R} \right)^{-\frac{1}{2}}
\en
We can evaluate $k$ in simple way: its definition is given by
Eq. (\ref{k}), which involves the integrals in
Eq. (\ref{nonnormalizzata2}) and in
Eq. (\ref{nonnormalizzata1}), thus:
\eqa
\label{calcolokparte1}
k=\frac{C_1}{C_2}=R \cdot
\frac{\int_0^L \!\!\! dt
\left( \dots \right)_{[t]^2}}{\int_0^R \!\!\! dz \int_0^L \!\!\! dt
\left( \dots \right)_{[z,t]^2}}
\ena
Since the two eigenfunction parts expressing the
$t$-dependence and $z$-dependence are factorized, the denominator
integrals factorize, too. Moreover, we know that the
$z$-dependent part of the eigenfunction is
expressed by Eq. (\ref{zdependence}); these observations allow
one to write:
\eq
\label{calcolokparte2}
k= R \frac{\int_0^L \!\!\! dt 
\left( \dots \right)_{[t]^2}}{ \int_0^R
\sin^2 \left( \frac{n\pi z}{R}
\right)
\int_0^L \!\!\! dt 
\left( \dots \right)_{[t]^2}} =
R \cdot \frac{1}{\frac{1}{2} \; R} = 2
\en
where we exploited the fact that $n \in \mathbf{N}_0$.\\
Thus, in the case when we allow for the presence of 
the boundary term, the first-order non-trivial 
contribution to the interquark
potential will be:
\eq\label{v1staresatto}
V_1^\star = 
\frac{1}{L} \log \eta \left( i \frac{L}{2R^\star} \right)
\en
It is important to stress the fact that this result does not
rely on the assumption to be in the ``large $\frac{L}{2R}$
regime'': as a matter of fact, the only hypothesis 
for the validity of Eq. (\ref{sostituzione}) is that
$b$ can be considered as a perturbatively small parameter.\\
On the other hand, if we restrict to the case
when $\frac{L}{2R} \gg 1$, then Eq. (\ref{v1staresatto})
can be approximated by:
\eqa\label{v1starapprossimato}
V_1^\star \!\!\!\!\!\!\!\!& & \simeq - \frac{\pi}{24 R^\star} = - \frac{\pi}{24 R \left( 1 + 
2 \frac{b}{R} \right)^{-\frac{1}{2}}} = \nonumber \\
& & = - \frac{\pi}{24 R} \sqrt{ 1 + 2 \frac{b}{R} }
\simeq - \frac{\pi}{24 R} \left( 1 + \frac{b}{R}  \right)
\ena
which exactly reproduces the term appearing in Eq. (3.9) in
\cite{lw02}.

\end{document}